\documentclass[preprint,12pt,numbers,sort&compress]{elsarticle}

\usepackage{longtable}
\usepackage[final]{microtype}    
\usepackage{amssymb}
\usepackage{bm}
\usepackage{booktabs}
\usepackage{amsmath}
\usepackage{microtype}
\usepackage{appendix}
\usepackage{graphicx}
\usepackage{tabularx}
\usepackage{colortbl}
\usepackage{xcolor}
\usepackage{makecell}

\usepackage{doi}

\usepackage{hyperref}
\hypersetup{
  colorlinks=true,
  linkcolor=blue,
  citecolor=blue,
  urlcolor=magenta,
  breaklinks=true,
  pdfauthor={Yonghao Wu, Chang Liu*, Vladimir Filaretov, Dmitry Yukhimets},
  pdftitle={Visual-Neural-Inspired Image Inpainting for Specific Objects-of-Interest Imaging},
}

\begin{document}

\begin{frontmatter}

\title{SIOI: A Lightweight Visual-Neural Network for Specific Objects-of-Interest Inpainting}

\author[1]{Yonghao Wu}
\author[2]{Chang Liu\corref{cor1}}
\author[3]{Vladimir Filaretov}
\author[3]{Dmitry Yukhimets}

\cortext[cor1]{Correspondence: liuchang@gdou.edu.cn}

\affiliation[1]{organization={School of Electronic Science and Engineering (School of Microelectronics), South China Normal University},
            addressline={Foshan, 528000},
            country={China}}
            
\affiliation[2]{organization={St. Petersburg School of Shipbuilding and Ocean Technology, Guangdong Ocean University},
            addressline={Zhanjiang 524088},
            country={China}}
            
\affiliation[3]{organization={Robotics Laboratory, Institute of Automation and Control Processes},
            addressline={Far Eastern Branch of the Russian Academy of Sciences},
            city={Vladivostok},
            postcode={690041},
            country={Russia}}

\begin{abstract}
Conventional image inpainting methods typically process entire images, which leads to computational inefficiency and background interference in object-centric tasks. Inspired by hierarchical cortical processing, we propose Specific Object-of-Interest Imaging (SIOI) — a lightweight, plug-and-play front end that extracts object-level structural and semantic priors to guide downstream inpainting. SIOI decouples object perception from texture synthesis. It produces background-suppressed object representations that significantly improve reconstruction quality and robustness. We evaluate SIOI on dedicated object datasets (Teapot, Elephant, Giraffe, Zebra) and demonstrate consistent improvements in SSIM, PSNR, MAE, and LPIPS across various backbone architectures and challenging conditions (low illumination, occlusion, motion blur). Theoretical analysis links our object-first strategy to attentional modulation in V1–V4, offering a biologically grounded rationale for target centric restoration. Results show that SIOI is an effective, efficient preprocessor that enhances object consistency and semantic coherence in modern inpainting pipelines.
\end{abstract}                                                                                                                                                                                                                                                                                       

\begin{graphicalabstract}
\end{graphicalabstract}

\begin{highlights}
\item Research highlight 1
\item Research highlight 2
\end{highlights}


\end{frontmatter}


\section{Introduction}
Image inpainting is a fundamental computer vision task that restores missing or corrupted regions of an image, such as those caused by occlusions, sensor noise, transmission errors, intentional editing, or physical damage. It is vital for applications such as cultural preservation, photography, medical imaging, visual effects, and autonomous systems. However, achieving accurate, semantically coherent inpainting remains a challenge.

Conventional inpainting methods are based primarily on intrinsic statistical properties and local image structures\citep{bertalmio2000image,bugeau2010comprehensive,guillemot2013image,yu2018generative,elharrouss2020image,qin2021image,xiang2023deep,yeh2017semantic}. Their core idea is to leverage information from intact neighboring regions to fill missing areas via propagation or patch-copying mechanisms. These methods work well for small, regular regions with regular texture, but struggle with large missing areas, complex boundaries, or semantically rich content. As a result, they often produce blurred textures, broken structures, and contextually inconsistent regions.

The rapid development of deep learning—particularly convolutional neural networks (CNNs), generative adversarial networks (GANs), and transformer- or diffusion-based models—has driven breakthroughs in pattern learning and high-fidelity image synthesis. These advances have catalyzed a paradigm shift in image inpainting\citep{yu2023inpaint,wang2025harmony,zhang2025sqsfill,liu2025transref,lu2025inpainting,mao2025aldii,liu2025nerf}. They capture rich priors and semantic representations from large datasets, enabling a more accurate reconstruction of missing content. Such models can reconstruct intricate details while maintaining overall plausibility.

Despite these advances, significant challenges remain. A key issue is inefficiency due to redundant information. In practice, inpainting is often object-centric rather than holistic. For example, in autonomous driving, medical imaging, and surveillance, over 90\% of visual data can be background. Whole-image models waste resources by analyzing irrelevant background, leading to inefficiency and reduced accuracy in reconstructing target objects. This calls for a paradigm shift from holistic processing to targeted object-focused inference.

Biological vision offers an efficient solution. Visual attention mechanisms prioritize relevant objects and suppress the background, enabling efficient recognition in complex scenes. This ``object-first'' strategy addresses information sparsity by first localizing targets and excluding background interference, improving focus and output quality.

Based on this insight, we propose Specific Object-of-Interest Imaging (SIOI). SIOI decouples object identification from the inpainting process. It functions as a front-end module that extracts clean object representations and suppresses background, enabling integration with any inpainting network. Our framework comprises two stages: object perception and object inpainting. By focusing the processing on the target regions, SIOI reduces background interference and improves both efficiency and accuracy. As shown in Table 1, this approach aligns with the hierarchy of the visual cortex, mimicking human attentional mechanisms.

Our main contributions are as follows:
\begin{enumerate}
    \item We propose SIOI, a novel framework that decouples object extraction from texture synthesis, thereby reducing background interference and generating useful structural priors for downstream tasks.
    \item We investigate connections between human visual attention and deep networks, integrating neuroscientific insights from visual cortical hierarchy to establish theoretical foundations.
    \item We demonstrate SIOI's effectiveness as a plug-and-play module through comprehensive experiments. Models equipped with SIOI show consistent improvements across metrics, datasets, and challenging scenarios.
\end{enumerate}

\begin{table}[htbp]
\centering
\small
\scriptsize 
\caption{Theoretical Framework Comparison:}
\label{tab:tab1}
\footnotesize
\begin{tabular}{
  >{\centering\arraybackslash}m{1.5cm}
  >{\centering\arraybackslash}m{2.9cm}
  >{\centering\arraybackslash}m{3.3cm}
  >{\centering\arraybackslash}m{1.8cm}
  >{\centering\arraybackslash}m{2.2cm}
}
\toprule
\textbf{Model} &
\textbf{Theoretical Framework} &
\textbf{Core Mechanism} &
\textbf{Processing Unit} &
\textbf{Biological Plausibility}\\
\midrule
Traditional &
Classical image processing&
Handcrafted features&
Entire image &
Not considered\\
\hline
Deep-Learning &
Deep neural networks&
Learned representations &
Entire image &
Not considered\\
\hline
\textbf{Ours} &
\textbf{Bio-inspired} &
\textbf{Object perception →~inpainting} &
\textbf{Specific Object-of-Interest} &
\textbf{V1-V4 simulation}\\
\bottomrule
\end{tabular}
\end{table}

\section{Related Works}
\label{sec:related_works}

Image inpainting has undergone multiple paradigm shifts, from early mathematical and patch-based models to modern deep generative architectures. Although these approaches differ in formulation and implementation, most state-of-the-art methods process the full image and therefore share a common ``whole-image'' bias.

\subsection{Traditional mathematical algorithms}
\label{subsec:traditional}
Early work emphasized low-level statistics, local geometric structure, and hand-made priors. Representative methods include partial-differential-equation (PDE) and diffusion-based models that propagate color and gradients while preserving edges \citep{bertalmio2000image,chan2001total,tschumperle2005vector}, and exemplar- or patch-based techniques that exploit non-local self-similarity by matching and copying patches from intact regions \citep{criminisi2004region,efros1999texture,wexler2007space}. These classical algorithms reconstruct missing areas via local propagation or patch synthesis \citep{10.1145/1201775.882267,hays2007scene,newson2014video}. They perform well on small, texture-regular regions, but typically fail for large occlusions or semantically rich content because they lack object-level understanding and global context.

\subsection{Data-driven neural approaches}
Modern data-driven methods learn priors from large datasets and substantially improve structural coherence and perceptual realism. The main paradigms include: (1) CNN encoder-decoder frameworks that exploit multi-scale features and attention mechanisms \citep{pathak2016context,yu2018generative,yu2020region,liu2020rethinking,kumar2023encoder}; (2) GAN-based approaches that use adversarial training to synthesize fine textures \citep{iizuka2017globally,wang2018image,fu2024text,sargsyan2023mi,liu2021pd,zhang2022gan,yu2022high}; (3) transformer-based methods that model long-range dependencies with self- and cross-attention \citep{yu2018generative,zhou2021transfill,li2022mat,deng2022t,dong2022incremental,shamsolmoali2023transinpaint,ko2023continuously}; and (4) diffusion models that recover details by iterative denoising \citep{lugmayr2022repaint,saharia2022palette,wei2023diffusion}.

Despite these gains, most neural methods still process entire images. This leads to three persistent issues for object-centric tasks: (i) information redundancy and inefficiency: processing background regions wastes computation and reduces effective information density (a notable problem in domains such as autonomous driving and satellite imaging); (ii) background interference: complex or ambiguous backgrounds can corrupt the reconstructed object structure and semantics when object and background share similar appearance; and (iii) lack of explicit object guidance: without explicit target specification, models struggle to separate target signals from background noise. Some methods incorporate user-provided cues (e.g., sketches or bounding boxes) to guide generation \citep{nazeri2019edgeconnect,9010689,liu2018image}, but such interactions are not always practical and often fail to achieve precise object separation in complex scenes.

\subsection{Biologically inspired insights and object-first strategies}
\label{subsec:bio_inspiration}
There is increasing evidence that hierarchical processing in biological vision correlates with the structure of modern neural networks \citep{cichy2016comparison,kok2013prior}. Human visual processing proceeds from the retina $\rightarrow$ LGN $\rightarrow$ V1 (edges) $\rightarrow$ V4 $\rightarrow$ IT (semantics) \citep{hubel1962receptive,felleman1991distributed}, which motivates multi-stage computational architectures \citep{ronneberger2015u,zhou2018unet++}. Predictive-coding theories \citep{friston2010free,lee2003hierarchical} further suggest that the brain uses priors to predict and correct sensory inputs, a concept analogous to the generator-discriminator dynamics in inpainting \citep{pathak2016context,yu2018generative,mirza2014conditional}.

Attention mechanisms—both bottom-up saliency and top-down task modulation—provide additional inspiration. Bottom-up processes highlight key features \citep{itti2001computational,miller2001integrative} and motivate saliency-weighted supervision \citep{wang2019saliencygan}; top-down control from the frontal regions motivates task-oriented modulation and conditional guidance. Edge-consistency losses and structural guidance \citep{nazeri2019edgeconnect} draw direct inspiration from early cortical responses to contours \citep{kanizsa1976subjective,kanizsa1985seeing,davis1994parallel}. Semantic guidance and multi-modal priors (e.g., CLIP-guided generation) demonstrate how higher-level signals can steer low-level synthesis \citep{saharia2022palette}. Memory-augmented retrieval mechanisms also resemble cortical-hippocampus recall \citep{goodfellow2020generative}.

Collectively, these biological principles motivate an ``object-first'' paradigm that prioritizes target extraction and suppresses irrelevant background. Our proposed Specific Object-of-Interest Imaging (SIOI) implementation implements this strategy as a plug-and-play front-end that amplifies object-relevant signals and reduces background interference, thereby addressing the limitations discussed above.

\section{Proposed Methodology}
\label{sec:methodology}

\begin{figure}[htbp]
   \centering
    \includegraphics[width=0.8\linewidth]{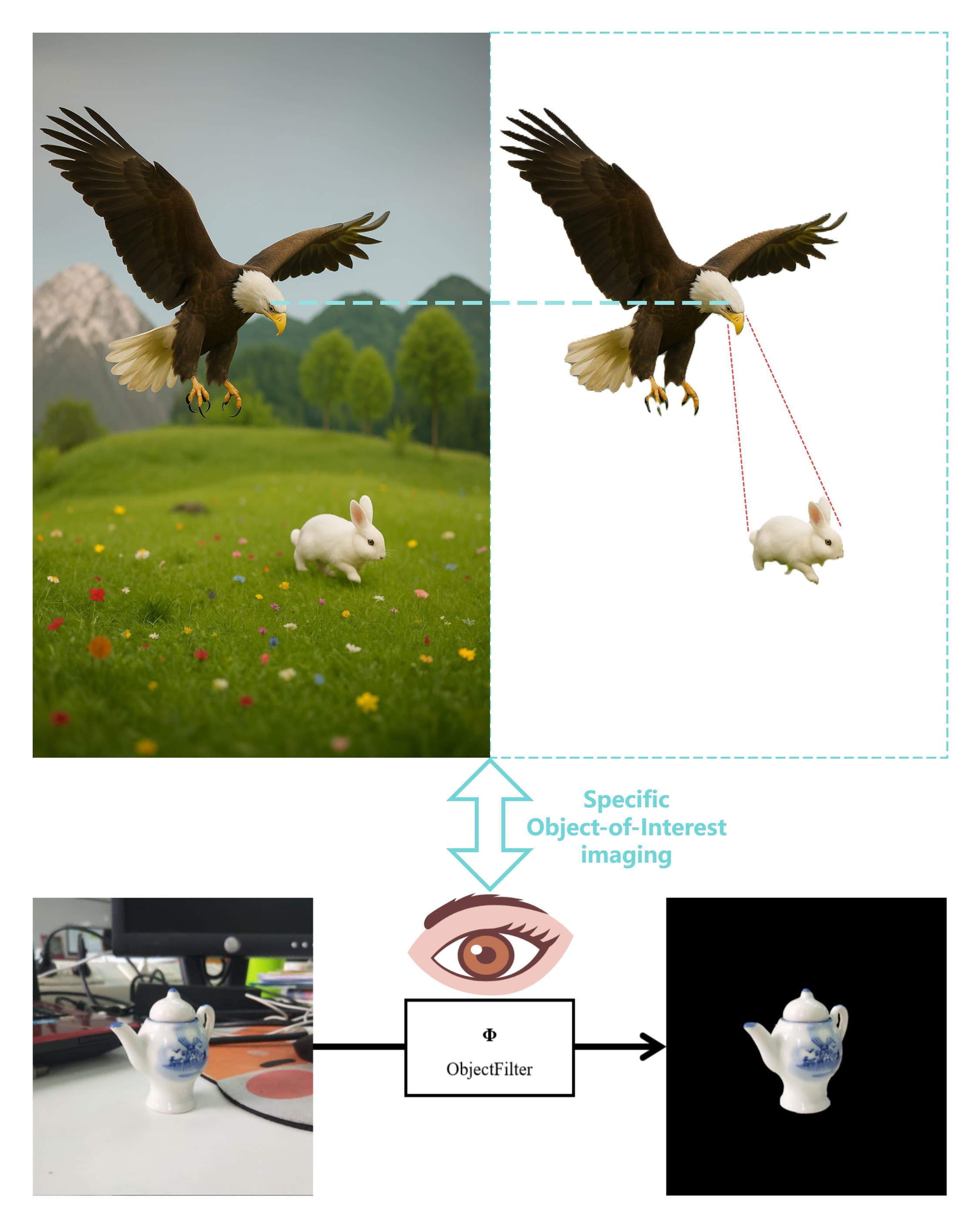}
    \caption{Specific Object-of-Interest Imaging (SIOI) pipeline.}
    \label{fig:fig1}
\end{figure}

This section presents a novel image inpainting framework that centers on object-focused reconstruction. Although conceptually related to previous object-centric techniques \citep{li2021optical,liao2022rgbd,chen2022object,li2017tile,wu2025detection}, our approach departs from them in key methodological aspects. The overall processing pipeline is illustrated in Fig.~\ref{fig:fig1}.

\subsection{Specific Object-of-Interest Imaging (SIOI)}
\begin{figure}[htbp]
   \centering
    \includegraphics[width=1\linewidth]{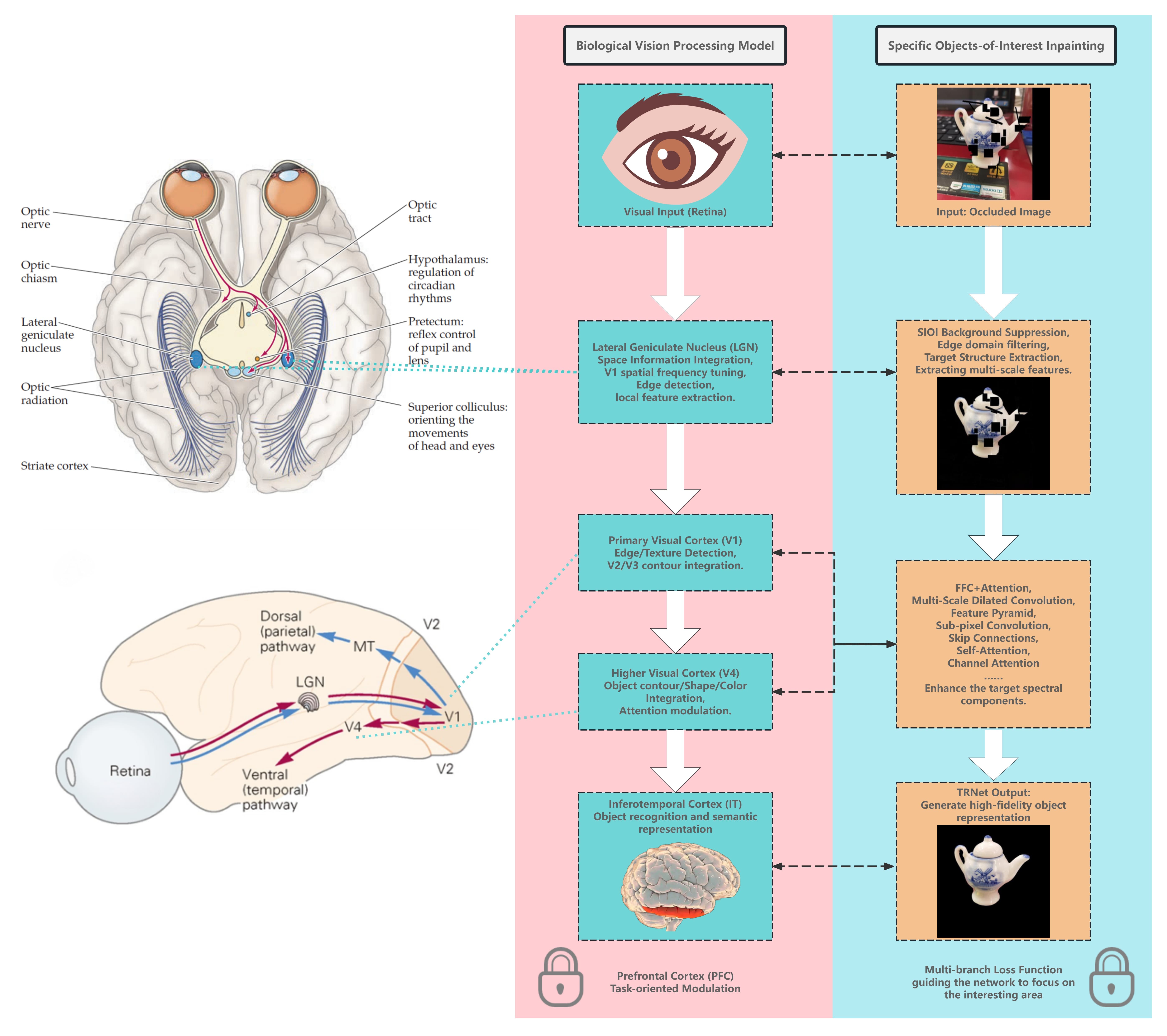}
    \caption{Proposed image inpainting framework based on SIOI.}
    \label{fig:fig2}
\end{figure}

SIOI, which we also refer to as attention-targeted imaging, is grounded in the neuroscientific principles of object-based attention. Desimone and Duncan's biased competition framework \cite{desimone1995neural} suggests that objects compete for representational resources in the visual cortex and that attention biases this competition to amplify targets and suppress background. fMRI evidence \cite{o1999fmri} further indicates that objects act as coherent attentional units and receive holistic enhancement. Additional studies show that object boundaries and task-driven factors modulate processing in early visual areas (V1–V4) and influence encoding resolution \citep{perry1984retinal,heywood1987role,serences2006voluntary,jigo2018attention,barbot2017attention,ekman2020object}. These findings motivate our design: a detection-guided saliency enhancement that applies higher-frequency emphasis to target regions, producing a biologically plausible prior for inpainting (see Fig.~\ref{fig:fig2}).

Anatomically, the ventral visual pathway processes information hierarchically: local feature extraction (retina $\to$ LGN) $\to$ orientation selectivity (V1) $\to$ more complex and holistic representations of objects (V4/IT). Motivated by this hierarchy, we treat a natural image \(k\) as a superposition of multiple components of latent objects.
\begin{equation}
k=\sum_{i=1}^{M} k_i,
\end{equation}
where \(k\in\mathbb{R}^{H\times W\times C}\) is the scene tensor, each \(k_i\in\mathbb{R}^{h_i\times w_i\times C}\) denotes the \(i\)-th object, and \(M\) is the total number of objects. The inpainting task emphasizes accurate separation and reconstruction of a target object \(k_\mu\) (\(\mu\in\{1,\dots,M\}\)) from cluttered backgrounds, similar to the perceptual “cocktail-party” effect.

To achieve robust separation under noise and occlusion, we introduce a bio-inspired dynamic feature modulation implemented through a target-sensitive, learnable measurement matrix \(\Gamma\). By differentiable learning, \(\Gamma\) yields multi-scale features \(\{\xi_j\}_{j=1}^N\) that span from simple edge/texture detectors (like V1) to complex shape encoders (like IT). For a target index \(\mu\), we express target measurements as:
\begin{equation}
\label{eq:target_feature_decomposition}
\Gamma k_\mu = \sum_{j=1}^{N} \omega_{j,\mu}\, \xi_j,
\end{equation}
where \(\omega_{j,\mu}\) are target-specific coefficients. When weights \(\omega_{j,\mu}\) focus on targets-relevant bases, scene-level measurements focus on the target:
\begin{equation}
\label{eq:scene_level_measurement}
\Gamma k \approx \widehat{k}_\mu,
\end{equation}
i.e., non-target contributions are suppressed when \(\eta_{j,m}\to\delta_{m\mu}\) for each channel \(j\).

Functionally, \(\Gamma\) implements an attentional gain analogous to modulation of V4, focusing on frequency bands and spatial regions important for target reconstruction. We also formalize object representations in a sparse dictionary \(\Psi\), so that each object admits a sparse expansion:
\begin{equation}
k_i = \sum_{j=1}^{M} \psi_{ij}\, x_{ij},
\end{equation}
with \(\psi_{ij}\) basis atoms and \(x_{ij}\) sparse coefficients. The scene then decomposes as
\begin{equation}
k = \sum_{i=0}^{N-1}\sum_{j=1}^{M} \psi_{ij}\, x_{ij},
\end{equation}
subject to sparsity constraints (\(\|\mathbf{x}_i\|_0 \ll M\)).

We implement these principles through a bio-inspired neural architecture: a Feedforward Feature Selection Network (FECNet) that emulates the ventral “what“ pathway to filter masked inputs \(M_h\), and a Target Reconstruction Network that simulates higher-level spatial integration. Training minimizes a reconstruction-based objective:
\begin{equation}
\{\xi, \theta\} = \arg\min_{\xi,\theta}\; \frac{1}{H}\sum_{h=1}^{H}\mathcal{C}\big(N_h,\tilde{N}_h\big),
\end{equation}
where \(\mathcal{C}(\cdot,\cdot)\) is an application-dependent loss (e.g., pixel, perceptual, and frequency-domain terms). After training, the learned filters \(\xi\) define a feature filter \(\phi\) such that
\begin{equation}
\phi k = \hat{k}_0,
\end{equation}
where \(\phi\) acts approximately as a projector onto the target subspace:
\begin{equation}
\phi(x_{ij}) =
\begin{cases}
x_{0j}, & \text{if } i = 0,\\
0, & \text{if } i \neq 0.
\end{cases}
\end{equation}

Under standard compressed sensing assumptions (sparsity, RIP on \(\Psi\), and bounded noise), \(\phi\) can be estimated in closed form by least-squares using the Moore–Penrose pseudoinverse:
\begin{equation}
\phi =
\Big(\sum_{h} \hat{k}_0^{(h)} \, k^{(h)\,T}\Big)
\Big(\sum_{h} k^{(h)} k^{(h)\,T}\Big)^\dagger,
\end{equation}
and are constrained as \(\phi \approx I - P_{\mathrm{bg}}\), where \(P_{\mathrm{bg}}\) project onto the background subspace. Equivalently, \(\phi\) solves
\begin{equation}
\label{eq:phi_best}
\phi^\star = \arg\min_{\phi\in\mathcal H}\; \big\|\,\phi - (I-P_{\mathrm{bg}})\,\big\|_F,
\end{equation}
in a Hilbert space \(\mathcal H\) endowed with the Frobenius norm \(\|\cdot\|_F\).

Our attention mechanism is implemented using the learned query and key vectors. Let \(\mathbf q_{xy}\) and \(\mathbf k_{ij}\) denote the query and key at positions \((x,y)\) and \((i,j)\), respectively, with the feature dimensionality \(d\). The attention weights follow the softmax-scaled inner product:
\begin{equation}
\label{eq:attn}
\alpha_{xy} = \frac{\exp\!\big(\langle \mathbf q_{xy},\mathbf k_{ij}\rangle / \sqrt{d}\big)}
{\sum_{i,j}\exp\!\big(\langle \mathbf q_{xy},\mathbf k_{ij}\rangle / \sqrt{d}\big)}.
\end{equation}
When queries and keys align, \(\alpha_{xy}\) approaches 1, generating attention gain consistent with Feature Integration Theory.

To ensure that the target is statistically distinguishable from background and other objects, we require
\begin{equation}
D_{\mathrm{KL}}\big(P_{\mathrm{target}}\|P_{\mathrm{background}}\big)
>
\sum_{i=2}^{M} D_{\mathrm{KL}}\big(P_{\mathrm{target}}\|P_{\mathrm{object}_i}\big),
\end{equation}
so that large-deviation arguments concentrate empirical distributions on entropy-minimizing configurations. Under Assumptions A1–A3 (object sparsity, RIP on \(\Psi\), bounded noise), the learned projector \(\phi\) approximates \(I-P_{\mathrm{bg}}\) and admits the error bound
\begin{equation}
\|\phi k - \widehat{k}_0\|_2 \le C_1\,\sigma_s(k) + C_2\,\epsilon,
\end{equation}
where \(C_1,C_2\) depend on the RIP constants and dictionary conditioning, \(\sigma_s(k)\) is the best \(s\)-term approximation error, and \(\epsilon\) bounds measurement noise.

These theoretical results provide a principled foundation for our object-centric imaging approach: formulating target separation as a sparsity-constrained projection and combining information-theoretic arguments with bio-inspired attention yields a system that, under the stated assumptions, isolates and reconstructs the target up to approximation and noise limits.

\subsection{Object-of-Interest Inpainting System}
\label{subsec:imaging_network}

\begin{figure}[htbp]
   \centering
    \includegraphics[width=1\linewidth]{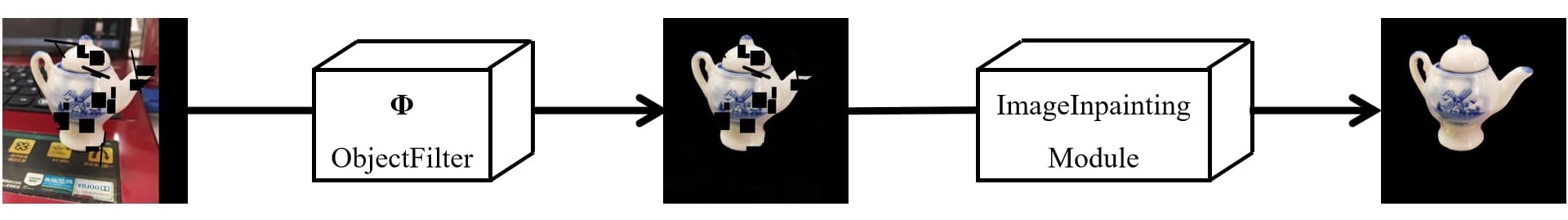}
    \caption{Processing flow of the proposed specific objects-of-interest inpainting system.}
    \label{fig:fig3}
\end{figure}

We propose a two-stage framework based on Specific Object-of-Interest Imaging (SIOI) to improve structural coherence and reconstruction accuracy for occluded images (Fig.~\ref{fig:fig3}). Unlike conventional pixel-wise RGB completion, our architecture adopts an object-first, biologically inspired strategy that decomposes the task into structure-prioritized refinements: (1) object perception and (2) object inpainting. The SIOI module directs the model to focus on target regions while suppressing background interference, thereby reducing computation and improving reconstruction fidelity.

\begin{figure}[htbp]
  \centering
  \includegraphics[width=\linewidth]{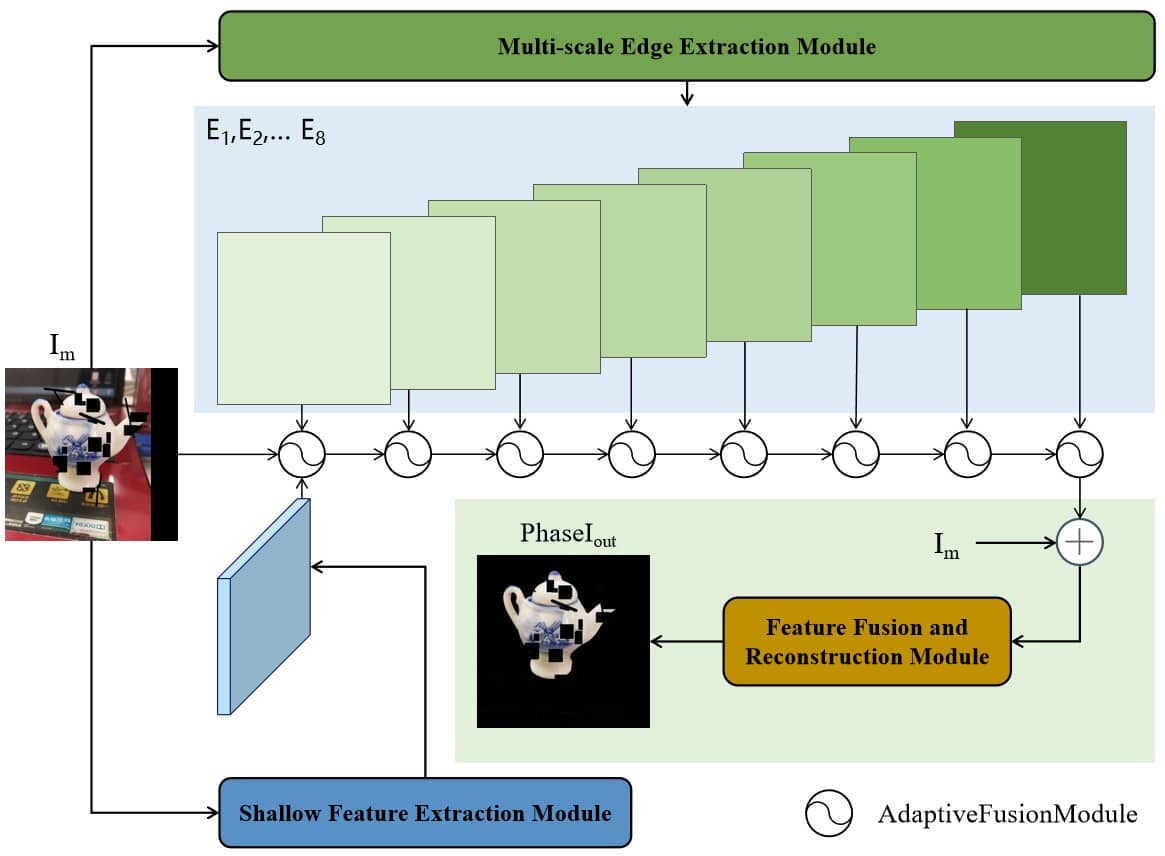}
  \caption{Architecture of the Specific Object-of-Interest Imaging (SIOI) module.}
  \label{fig:fig4}
\end{figure}

\textbf{Stage I: Specific Object-of-Interest Imaging.}Biological vision begins with retinal phototransduction and thalamic preprocessing. The lateral geniculate nucleus (LGN) integrates spatial input and attenuates high-frequency noise through center–surround antagonism, while the primary visual cortex (V1) extracts oriented edges and spatial-frequency textures \citep{hubel1962receptive}. Inspired by this cascade, the SIOI module produces structurally coherent object-centric priors from complex or occluded scenes (Fig.~\ref{fig:fig4}). Practically, SIOI is implemented as an Enhanced Feature-Selective Large-Kernel Network tailored for structural prior generation and restoration.

\begin{figure}[htbp]
  \centering
  \includegraphics[width=\linewidth]{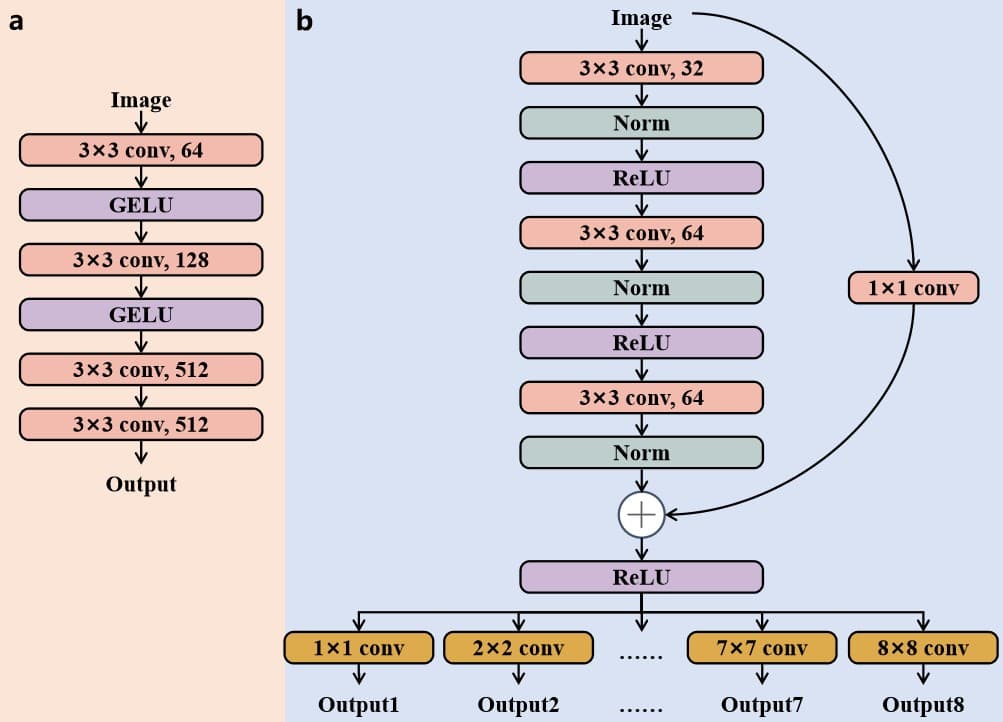}
  \caption{(a) Shallow feature extraction module; (b) Multi-scale edge extraction module.}
  \label{fig:fig5}
\end{figure}
\begin{figure}[htbp]
  \centering
  \includegraphics[width=0.7\linewidth]{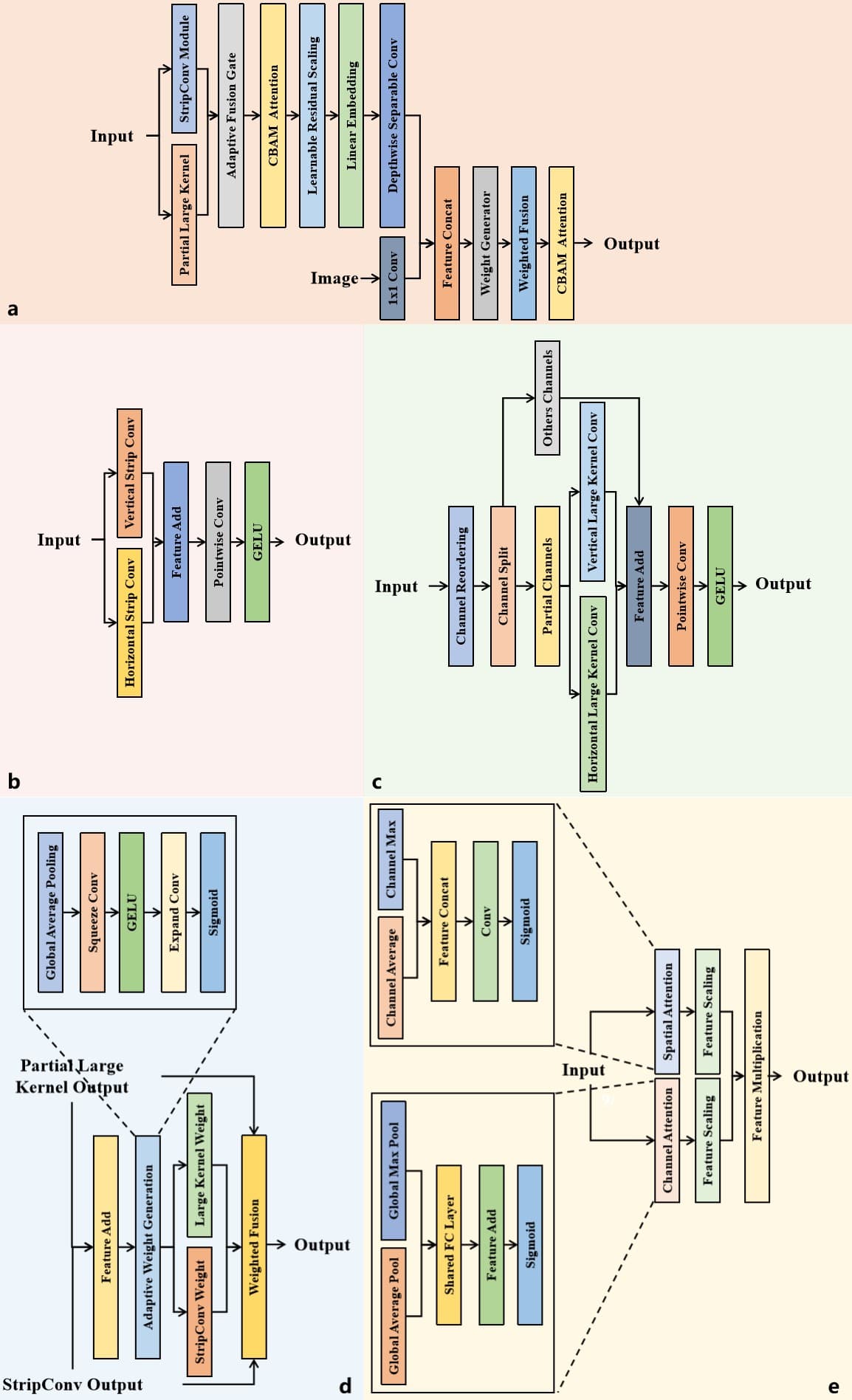}
  \caption{(a) Feature fusion and reconstruction module; (b) StripConv module; (c) Partial large-kernel branch; (d) Adaptive fusion gate; (e) CBAM attention.}
  \label{fig:fig6}
\end{figure}

The proposed SIOI pipeline integrates four key components: a shallow feature extractor, a Multi-scale Edge Extraction Module, a cascade of Adaptive Fusion Modules—each coupled with an Enhanced Edge Information Fusion (EIF) unit—and a reconstruction head. The Multi-scale Edge Extraction Module adopts a dilation-pyramid strategy to generate multi-receptive-field edge maps, effectively capturing both fine-grained details and coarse structural information, as illustrated in Fig.~\ref{fig:fig5}. Each Adaptive Fusion Module consists of two complementary branches: a Strip Convolution branch, which utilizes depthwise-separable convolutions with large rectangular kernels (e.g. $31\times1$ and $1\times31$) to efficiently model long-range directional dependencies, and a Partial Large-Kernel branch, where large-kernel processing is selectively applied to a subset of channels (default 25\%) in conjunction with channel shuffling to facilitate cross-channel interactions at minimal computational cost. The outputs of both branches are adaptively fused via a learnable gating mechanism that dynamically balances their contributions, as shown in Fig.~\ref{fig:fig6}.

EIF units inject multi-scale edge features into Adaptive Fusion outputs in a hierarchical, cascaded manner, which preserves structural consistency and reduces geometric distortion during deep processing. Convolutional Block Attention Modules (CBAM) are integrated at selected stages to model channel and spatial attention jointly, emphasizing semantically relevant regions while suppressing background clutter. The reconstruction path performs progressive refinement using global residual connections and adaptive residual scaling to stabilize training and retain low-frequency content.

Formally, let \(I_{\mathrm{in}}\) denote the input image, \(F_{\mathrm{shallow}}\) the shallow features, and \(Edge_i(I_{\mathrm{in}})\) the multi-scale edge maps in stage \(i\). Denote by \(\mathrm{AFM}_i(\cdot)\) the Adaptive Fusion Module in stage \(i\), by \(\mathrm{EIF}_i(\cdot,\cdot)\) the corresponding edge-fusion operator, and by \(R(\cdot)\) the reconstruction operator. Then
\begin{equation}
  I_{\mathrm{out}} \;=\; R\Bigg(\; \sum_{i=1}^{N} \mathrm{EIF}_i\big( \mathrm{AFM}_i(F_{\mathrm{shallow}}),\; Edge_i(I_{\mathrm{in}}) \big) \;+\; F_{\mathrm{shallow}} \Bigg).
  \label{eq:overall}
\end{equation}
Equation~\eqref{eq:overall} highlights the hierarchical fusion of adaptive processing with multi-scale edge priors, followed by reconstruction that combines deep refinements with shallow cues.

Functionally, Stage I implements V1-like edge extraction and extrastriate-like attentional gain to produce object-prior maps that emphasize salient targets and suppress background noise \citep{desimone1995neural,ekman2023successor}. The multi-scale edge guidance reduces structural distortion; the dual-path large-kernel design captures directional and global contexts; selective channel processing and depthwise separable convolutions maintain efficiency; attention modules increase selectivity; and hierarchical fusion preserves structural fidelity across depths. Together, these components transform complex, occluded scenes into structurally coherent, object-centered representations that provide strong priors for subsequent texture synthesis.

\textbf{Stage II: Specific Objects-of-Interest Inpainting.}
To independently validate the efficacy of the SIOI priors, Stage II employs a deliberately simple inpainting network that does not require manual masks. This design isolates the contribution of SIOI: if high-quality reconstructions are obtained with a modest backbone, the improvement can be attributed to the priors rather than to an over-parameterized inpainting model.

\begin{figure}[htbp]
\centering
\includegraphics[width=1\linewidth]{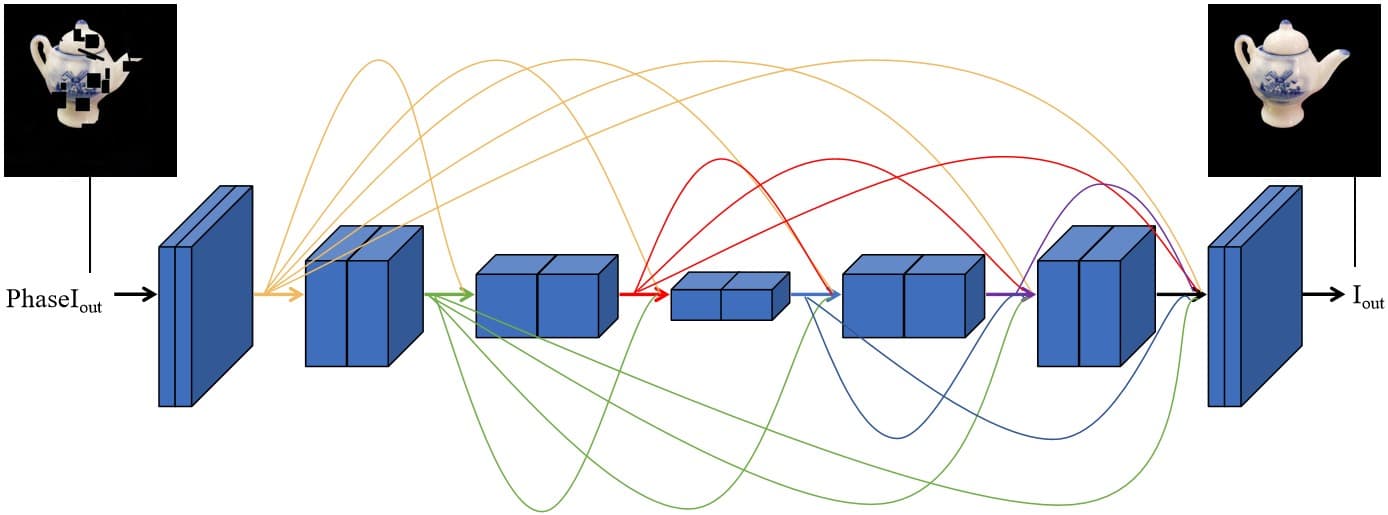}
\caption{Stage II inpainting architecture: ResNet encoder–decoder for structure and SRCNN branch for detail enhancement.}
\label{fig:fig7}
\end{figure}

The inpainting network is built on a symmetric encoder–decoder architecture with a ResNet backbone\citep{he2016deep}, which is responsible for structural reconstruction, combined with a lightweight SRCNN-style module\citep{dong2015image} dedicated to high-frequency detail enhancement (Fig.~\ref{fig:fig7}). Key design elements include residual blocks in both the encoder and decoder to capture and propagate structural information, skip connections that preserve multi-scale spatial features, and an SRCNN-based detail branch to refine local textures and sharpness. This hybrid strategy effectively allocates structural estimation and semantic guidance to the ResNet backbone, while delegating fine-detail synthesis to the SRCNN module. As a result, the SIOI component demonstrates its versatility as a plug-and-play unit that adapts to diverse inpainting pipelines.

\subsection{Loss function}
To produce high-fidelity reconstructions, we adopt a multi-branch composite loss that jointly enforces structural coherence, perceptual fidelity, and texture/style consistency. The total loss integrates: (1) a spatially weighted reconstruction loss, (2) a perceptual loss, and (3) a style loss. Appropriate weighting and multi-stage optimization balance these objectives.

\textbf{Weighted Reconstruction Loss.} A pixel-wise weight map \(w(x,y)\), derived from target priors, biases reconstruction errors toward semantically important regions:
\begin{equation}
\mathcal{L}_{\mathrm{weighted}} \;=\; \frac{1}{N}\sum_{x,y}\big(\alpha\,w(x,y)+1\big)\,\big\|\hat{I}(x,y)-I_{\mathrm{gt}}(x,y)\big\|_2^2,
\end{equation}
or equivalently,
\begin{equation}
\mathcal{L}_{\mathrm{recon}} \;=\; \sum_i w_i\,\big\|I_i^{\mathrm{pred}}-I_i^{\mathrm{gt}}\big\|_2^2,
\end{equation}
where \(\alpha\) (empirically set to 4) controls the emphasis on target regions. This weighting reduces edge blurring and structural discontinuities by focusing supervision on critical areas.

\textbf{Perceptual Loss.} To capture high-level semantics, we use a VGG-16-based perceptual loss:
\begin{equation}
\mathcal{L}_{\mathrm{perc}} \;=\; \sum_{l\in\mathcal L} \frac{1}{C_lH_lW_l}\,\big\|\phi_l(I^{\mathrm{pred}})-\phi_l(I^{\mathrm{gt}})\big\|_2^2,
\end{equation}
where \(\phi_l(\cdot)\) are the feature maps of the selected VGG-16 layers.

\textbf{Style Loss.} Texture and style consistency are enforced via Gram-matrix statistics:
\begin{equation}
G_l(x) \;=\; \phi_l(x)\,\phi_l(x)^\top,\qquad
\mathcal{L}_{\mathrm{style}} \;=\; \sum_{l\in\mathcal L}\big\|G_l(I^{\mathrm{pred}})-G_l(I^{\mathrm{gt}})\big\|_F^2.
\end{equation}

\textbf{Composite Loss.} The final objective is
\begin{equation}
\mathcal{L}_{\mathrm{total}} = \lambda_1 \mathcal{L}_{\mathrm{recon}} + \lambda_2 \mathcal{L}_{\mathrm{perc}} + \lambda_3 \mathcal{L}_{\mathrm{style}},
\end{equation}
with \(\lambda_1=1.0\), \(\lambda_2=0.1\), and \(\lambda_3=250\). The reconstruction term enforces edge fidelity in key regions, the perceptual term preserves semantic structure, and the style term mitigates texture artifacts.

\section{Experiment}
\label{sec:experiment}

\subsection{Datasets}

We evaluated our method on four datasets. First, we built an internal object-centric \emph{Teapot} dataset comprising \(2{,}000\) training samples and \(200\) test samples. Each sample is a \(512\times512\) RGB image of a teapot captured from multiple viewpoints in cluttered indoor scenes; each image is paired with a carefully annotated \(512\times512\) ground-truth mask. Masks were produced via a multi-stage manual labeling and verification pipeline to ensure accurate geometric and textural alignment with object boundaries.

In addition, we derive three object-specific subsets from COCO2017: \emph{elephant} (2{,}143 training, 89 validation/test samples), \emph{giraffe} (2{,}546 training, 101 validation/test samples) and \emph{zebra} (1{,}916 training, 85 validation/test samples). All images were resized to \(512\times512\) for consistency.

For all datasets, we apply a uniform occlusion protocol: occlusion masks are automatically generated from object maps so that they cover only foreground objects, occluding roughly \(\sim40\%\) of the object area. This setting preserves contextual cues while requiring non-trivial structural reconstruction during training and inference.

\subsection{Experimental setup}




All experiments were conducted on Ubuntu 20.04 using an Intel Xeon Platinum 8358P CPU and an NVIDIA GeForce RTX 4090 GPU (24 \,GB VRAM). The framework was implemented in PyTorch 1.10.0 with CUDA 11.3. Training was carried out for 50 epochs with a batch size of 4 (limited by GPU memory), using Adam optimizer and an initial learning rate of $2\times10^{-4}$. We followed a two-stage training protocol. In Stage I (pretraining), the Specific Objects-of-Interest Imaging (SIOI) module was pretrained using only reconstruction loss $\mathcal{L}_{\mathrm{recon}}$ to learn intermediate structurally informative representations; after pretraining, the SIOI module was frozen and kept fixed. In Stage II (fine-tuning), with the SIOI module frozen, the inpainting network was fine-tuned using perceptual loss $\mathcal{L}_{\mathrm{perc}}$ and style loss $\mathcal{L}_{\mathrm{style}}$ to improve semantic coherence and texture fidelity. The resulting SIOI network has a size of approximately 10 MB. The source code is publicly available at: \url{https://github.com/WYH302/WuYonghao}.

\textbf{Baselines}
We compare against a representative set of inpainting methods spanning exemplar-based, patch-based, GAN-, transformer- and diffusion-style approaches. These include Criminisi~\citep{criminisi2004region}, PatchMatch~\citep{barnes2009patchmatch}, Edge-Connect~\citep{nazeri2019edgeconnect}, MISF~\citep{li2022misf}, AOT-GAN~\citep{zeng2022aggregated}, LaMa~\citep{suvorov2022resolution}, and RePaint~\citep{lugmayr2022repaint}. All baselines were run for five independent trials with random initializations, using official author implementations and pretrained weights, where available.

\textbf{Evaluation metrics}
We report four standard metrics: structural similarity (SSIM, higher is better), peak signal-to-noise ratio (PSNR, higher is better), mean absolute error (MAE, lower is better), and Learned Perceptual Image Patch Similarity (LPIPS) \citep{zhang2018unreasonable} (lower is better).

\subsection{Preliminary validation of SIOI (controlled experiment)}

To isolate the contribution of the SIOI module, we performed a controlled experiment using a deliberately simple Stage II inpainting network. The intent is to demonstrate that the quality of reconstructions predominantly arises from SIOI's structural and semantic priors rather than from an over-parameterized decoder.

\begin{figure}[htbp]
\centering
\includegraphics[width=1\linewidth]{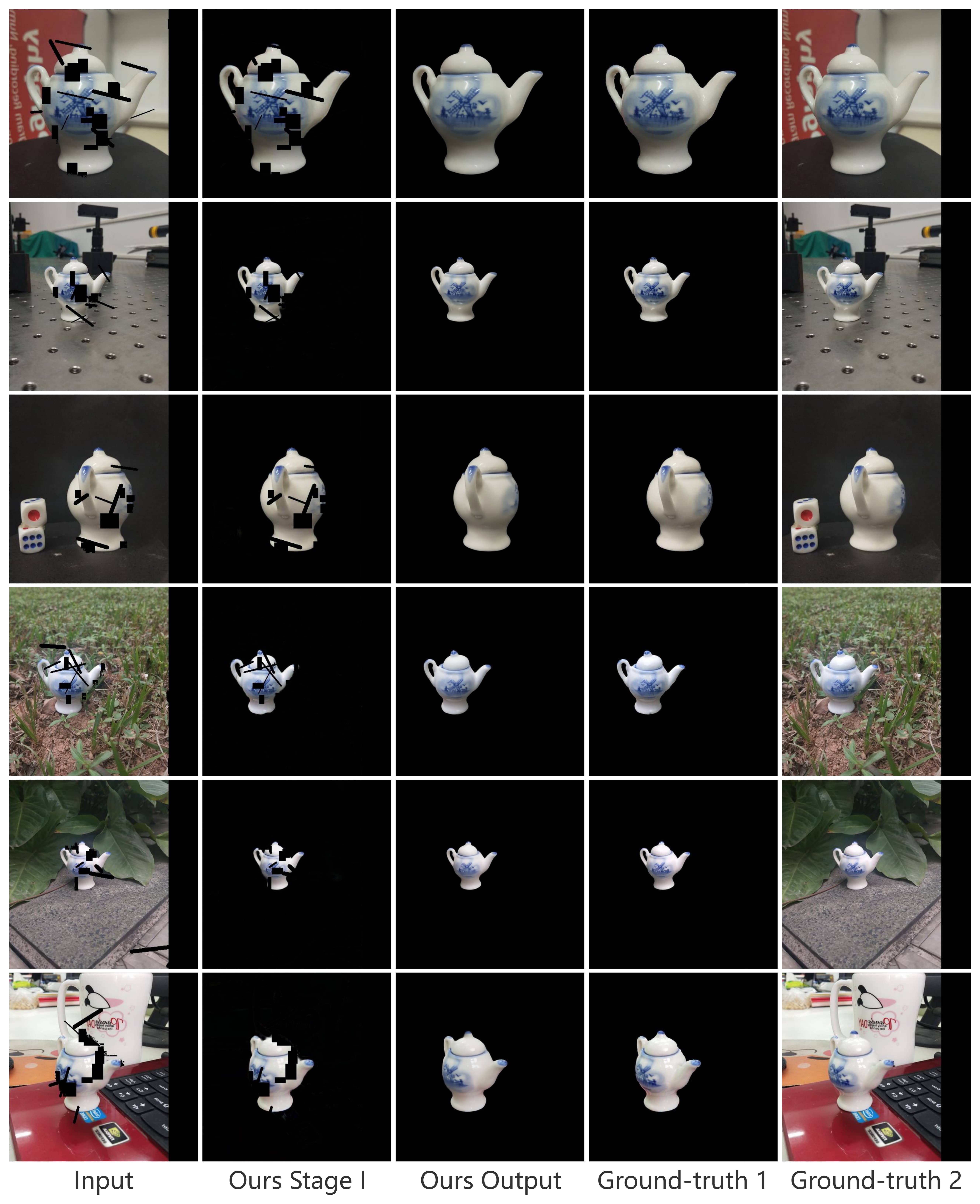}
\caption{Representative reconstructions from the preliminary validation on the Teapot dataset.}
\label{fig:fi8}
\end{figure}

Our model automatically identifies and inpaints corrupted regions without manual mask annotations. To avoid trivial cues, we retained black borders in the inputs so that the network could not rely solely on color artifacts for corruption detection. The results of the Teapot dataset are summarized in Table~\ref{tab:tab2}, and example reconstructions are shown in Fig.~\ref{fig:fi8}. The high SSIM/PSNR and low LPIPS values indicate that SIOI produces clean, structurally coherent object representations that substantially ease downstream inpainting.

\begin{table}[htbp]
\centering
\small
\caption{Preliminary validation on the Teapot dataset (mean \(\pm\) std over five runs). Arrows indicate preferred direction.}
\label{tab:tab2}
\begin{tabular}{ccccc}
\toprule
Metric & SSIM \(\uparrow\) & PSNR (dB) \(\uparrow\) & MAE \(\downarrow\) & LPIPS \(\downarrow\) \\
\midrule
Experimental values & \(0.98 \pm 0.02\) & \(33.86 \pm 0.38\) & \(1.61 \pm 0.18\) & \(0.02 \pm 0.01\) \\
\bottomrule
\end{tabular}
\end{table}

\subsection{Generalized empirical evaluation — quantitative comparisons}

We further evaluate SIOI as a plug-and-play front-end: for each baseline method, we prepend SIOI and measure performance under a unified object-level evaluation protocol (outputs are multiplied by the corresponding object masks). Table~\ref{tab:tab3} reports numerical results (mean over five runs). Across datasets and metrics, SIOI consistently improves downstream performance. The principal observation is that significant gains are achieved by providing purified, object-centric priors to diverse inpainting pipelines, rather than by modifying the downstream model architectures.

For example, when combined with classical exemplar-based methods, SIOI significantly improves PSNR (e.g., Criminisi + SIOI yields a \(\sim7\) dB PSNR gain under our protocol). When paired with modern spectral/diffusion methods (e.g., LaMa), SIOI further enhances both PSNR and LPIPS, demonstrating that object-centric preprocessing complements both the deterministic and stochastic reconstruction paradigms. Improvements are most pronounced on the Teapot dataset, where SIOI+LaMa achieves a near-perfect reconstruction (SSIM: 0.9973, LPIPS: 0.0023). These results corroborate that SIOI (i) facilitates target separation across complex backgrounds and (ii) reduces redundant processing by focusing computation on relevant regions.

\begin{table}[htbp]
\small
\scriptsize 
\renewcommand{\arraystretch}{0.8} 
\centering
\caption{Experimental Results for Different Datasets and Models, ↑ means higher is better, and ↓ means lower is better.}
\label{tab:tab3}
\begin{tabular}{clcccc}
\toprule
\textbf{Datasets} & \textbf{Model} & \textbf{SSIM} $\uparrow$ & \textbf{PSNR(dB)} $\uparrow$ & \textbf{MAE} $\downarrow$ & \textbf{LPIPS} $\downarrow$ \\
\midrule
{\textbf{Teapot}} 
 & Criminisi & 0.9415 & 24.597 & 3.5517 & 0.0965 \\
 & PatchMatch & 0.9212 & 21.219 & 5.4619 & 0.1263 \\
 & Edge-Connect & 0.9457 & 25.095 & 3.3687 & 0.0908 \\
 & AOT-GAN & 0.9487 & 25.329 & 3.2564 & 0.0863 \\
 & MISF & 0.9370 & 24.871 & 3.7130 & 0.1178 \\
 & LaMa & 0.9524 & 25.534 & 3.1024 & 0.0833 \\
 & RePaint & 0.9330 & 24.037 & 4.1087 & 0.1141 \\
  & SIOI+Criminisi & 0.9841(↑4.5\%) & 31.5520(↑28.3\%) & 0.7396(↓79.2\%) & 0.0246(↓74.5\%) \\
 & SIOI+PatchMatch & 0.9752(↑5.9\%) & 25.9120(↑22.1\%) & 1.7894(↓67.2\%) & 0.0408(↓67.7\%) \\
 & SIOI+Edge-Connect & 0.9795(↑3.6\%) & 32.3780(↑29.0\%) & 0.8364(↓75.2\%) & 0.0299(↓67.1\%) \\
 & SIOI+AOT-GAN & 0.9248(↓2.5\%) & 36.9400(↑45.8\%) & 1.0471(↓67.8\%) & 0.0094(↓89.1\%) \\
 & SIOI+MISF & 0.9739(↑3.9\%) & 33.4870(↑34.6\%) & 1.0541(↓71.6\%) & 0.0674(↓42.8\%) \\
 & SIOI+LaMa & 0.9973(↑4.7\%) & 45.2220(↑77.1\%) & 0.1119(↓96.4\%) & 0.0023(↓97.2\%) \\
 & SIOI+RePaint & 0.9712(↑4.1\%) & 28.5990(↑19.0\%) & 1.6907(↓58.9\%) & 0.0645(↓43.5\%) \\
\midrule
{\textbf{elephant}} 
 & Criminisi & 0.9671 & 32.615 & 1.3453 & 0.0262 \\
 & PatchMatch & 0.9596 & 28.643 & 2.1005 & 0.0441 \\
 & Edge-Connect & 0.9562 & 31.958 & 1.7455 & 0.0423 \\
 & AOT-GAN & 0.8416 & 27.653 & 4.9353 & 0.0608 \\
 & MISF & 0.8472 & 23.789 & 5.2913 & 0.1908 \\
 & LaMa & 0.9236 & 26.059 & 2.9596 & 0.0949 \\
 & RePaint & 0.8624 & 23.690 & 5.1769 & 0.1676 \\
 & SIOI+Criminisi & 0.9774(↑1.1\%) & 33.1520(↑1.6\%) & 0.8312(↓38.2\%) & 0.0190(↓27.5\%) \\
 & SIOI+PatchMatch & 0.9721(↑1.3\%) & 29.3270(↑2.4\%) & 1.3736(↓34.6\%) & 0.0325(↓26.3\%) \\
 & SIOI+Edge-Connect & 0.9636(↑0.8\%) & 30.6930(↓4.0\%) & 1.2479(↓28.5\%) & 0.0448(↑5.9\%) \\
 & SIOI+AOT-GAN & 0.9341(↑11.0\%) & 34.6910(↑25.5\%) & 1.4350(↓70.9\%) & 0.0243(↓60.0\%) \\
 & SIOI+MISF & 0.8883(↑4.9\%) & 29.0700(↑22.2\%) & 3.4237(↓35.3\%) & 0.1848(↓3.1\%) \\
 & SIOI+LaMa & 0.9786(↑6.0\%) & 34.7060(↑33.2\%) & 0.6523(↓78.0\%) & 0.0140(↓85.2\%) \\
 & SIOI+RePaint & 0.9102(↑5.5\%) & 28.9560(↑22.2\%) & 3.2348(↓37.5\%) & 0.1489(↓11.2\%) \\
\midrule
{\textbf{giraffe}} 
 & Criminisi & 0.9721 & 29.962 & 1.5050 & 0.0184 \\
 & PatchMatch & 0.9653 & 27.671 & 2.3031 & 0.0368 \\
 & Edge-Connect & 0.9678 & 29.914 & 1.6754 & 0.0280 \\
 & AOT-GAN & 0.9109 & 26.742 & 4.1489 & 0.0388 \\
 & MISF & 0.8998 & 23.505 & 4.6733 & 0.1208 \\
 & LaMa & 0.9424 & 25.732 & 2.5901 & 0.0670 \\
 & RePaint & 0.9079 & 23.563 & 4.5075 & 0.1096 \\
 & SIOI+Criminisi & 0.9788(↑0.7\%) & 29.8240(↓0.5\%) & 1.1000(↓26.9\%) & 0.0170(↓7.6\%) \\
 & SIOI+PatchMatch & 0.9734(↑0.8\%) & 26.3270(↓4.9\%) & 1.7493(↓24.0\%) & 0.0306(↓16.8\%) \\
 & SIOI+Edge-Connect & 0.9676(↑0.0\%) & 27.5270(↓8.0\%) & 1.6295(↓2.7\%) & 0.0382(↑36.4\%) \\
 & SIOI+AOT-GAN & 0.8741(↓4.0\%) & 30.8610(↑15.4\%) & 2.3573(↓43.2\%) & 0.0350(↓9.8\%) \\
 & SIOI+MISF & 0.9325(↑3.6\%) & 27.2700(↑16.0\%) & 3.0461(↓34.8\%) & 0.1215(↑0.6\%) \\
 & SIOI+LaMa & 0.9796(↑3.9\%) & 31.0230(↑20.6\%) & 0.9491(↓63.4\%) & 0.0129(↓80.7\%) \\
 & SIOI+RePaint & 0.9465(↑4.3\%) & 27.1090(↑15.0\%) & 2.8830(↓36.0\%) & 0.0992(↓9.5\%) \\
\midrule
{\textbf{zebra}} 
 & Criminisi & 0.9677 & 27.598 & 2.2333 & 0.0236 \\
 & PatchMatch & 0.9262 & 22.627 & 4.1497 & 0.0857 \\
 & Edge-Connect & 0.9588 & 27.891 & 3.1040 & 0.0263 \\
 & AOT-GAN & 0.8451 & 22.399 & 9.6206 & 0.0486 \\
 & MISF & 0.8583 & 21.409 & 8.2618 & 0.1596 \\
 & LaMa & 0.9331 & 24.284 & 3.7468 & 0.0684 \\
 & RePaint & 0.8677 & 21.544 & 8.0205 & 0.1499 \\
 & SIOI+Criminisi & 0.9750(↑0.8\%) & 27.3890(↓0.8\%) & 1.6940(↓24.1\%) & 0.0211(↓10.6\%) \\
 & SIOI+PatchMatch & 0.9675(↑4.5\%) & 24.6910(↑9.1\%) & 2.4553(↓40.8\%) & 0.0365(↓57.4\%) \\
 & SIOI+Edge-Connect & 0.9632(↑0.5\%) & 26.0890(↓6.5\%) & 2.2360(↓28.0\%) & 0.0387(↑47.1\%) \\
 & SIOI+AOT-GAN & 0.9070(↑7.3\%) & 28.0250(↑25.1\%) & 3.2725(↓66.0\%) & 0.0527(↑8.4\%) \\
 & SIOI+MISF & 0.8960(↑4.4\%) & 24.0480(↑12.3\%) & 6.2437(↓24.4\%) & 0.1556(↓2.5\%) \\
 & SIOI+LaMa & 0.9772(↑4.7\%) & 28.8110(↑18.6\%) & 1.3896(↓62.9\%) & 0.0144(↓78.9\%) \\
 & SIOI+RePaint & 0.9130(↑5.2\%) & 24.3090(↑12.8\%) & 5.9439(↓25.9\%) & 0.1365(↓8.9\%) \\
\bottomrule
\end{tabular}
\end{table}

\subsection{Generalized empirical evaluation — Qualitative comparisons}

\begin{figure}[htbp]
\centering
\includegraphics[width=1\linewidth]{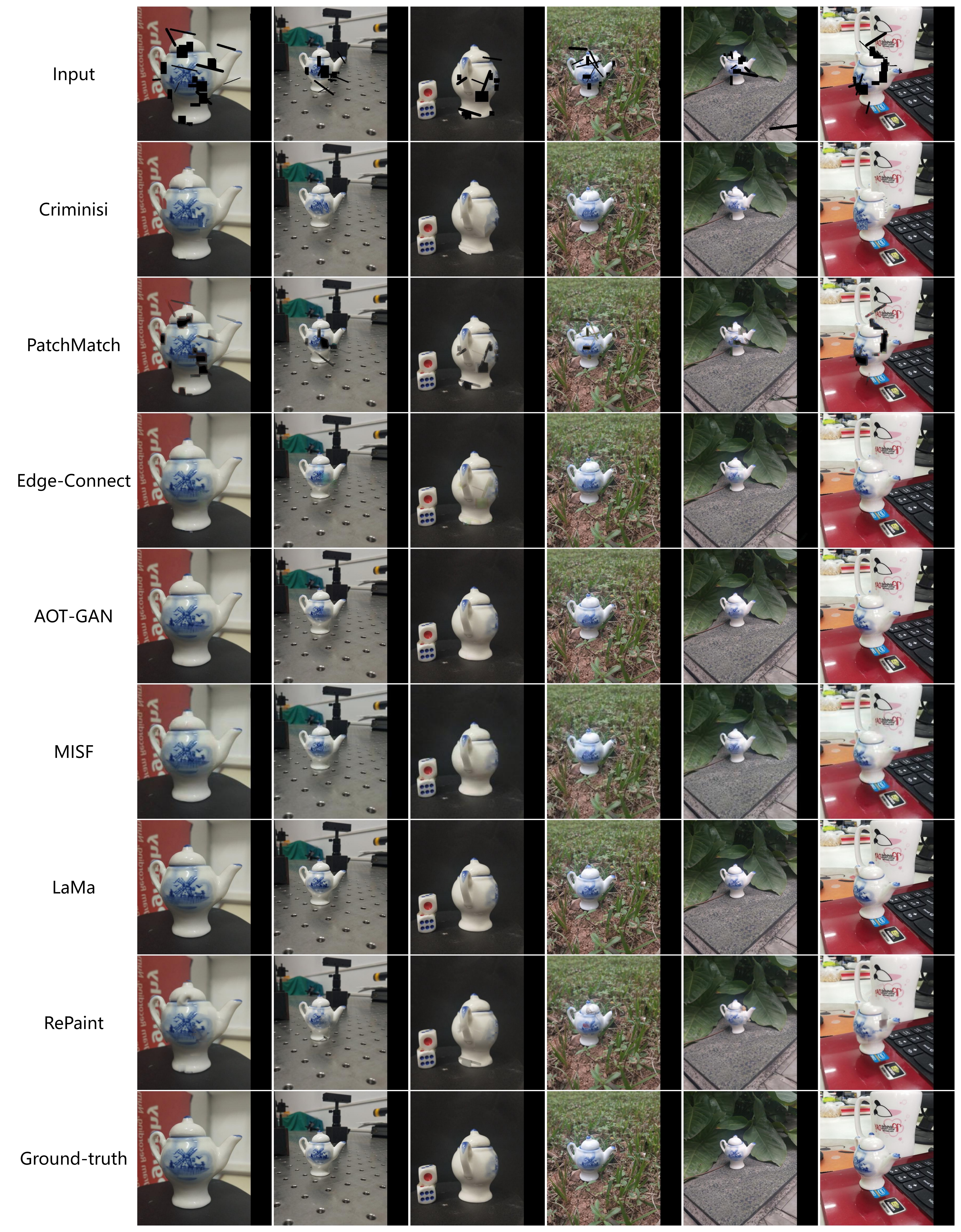}
\caption{Qualitative comparison of exemplar inpainting results (teapot dataset).}
\label{fig:fig9}
\end{figure}

\begin{figure}[htbp]
\centering
\includegraphics[width=1\linewidth]{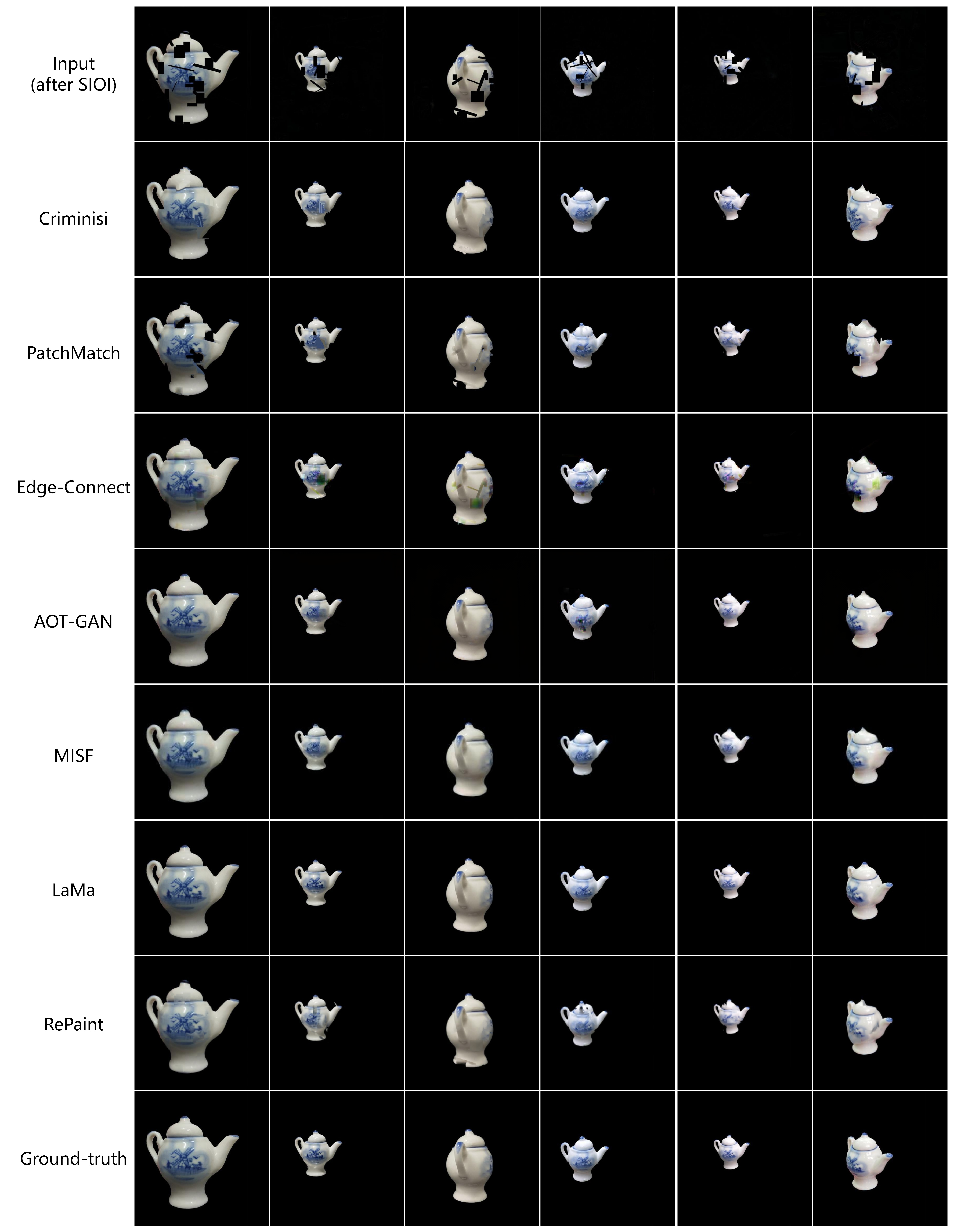}
\caption{Qualitative comparison of exemplar SIOI inpainting results (teapot dataset).}
\label{fig:fig10}
\end{figure}

\begin{figure}[htbp]
\centering
\includegraphics[width=1\linewidth]{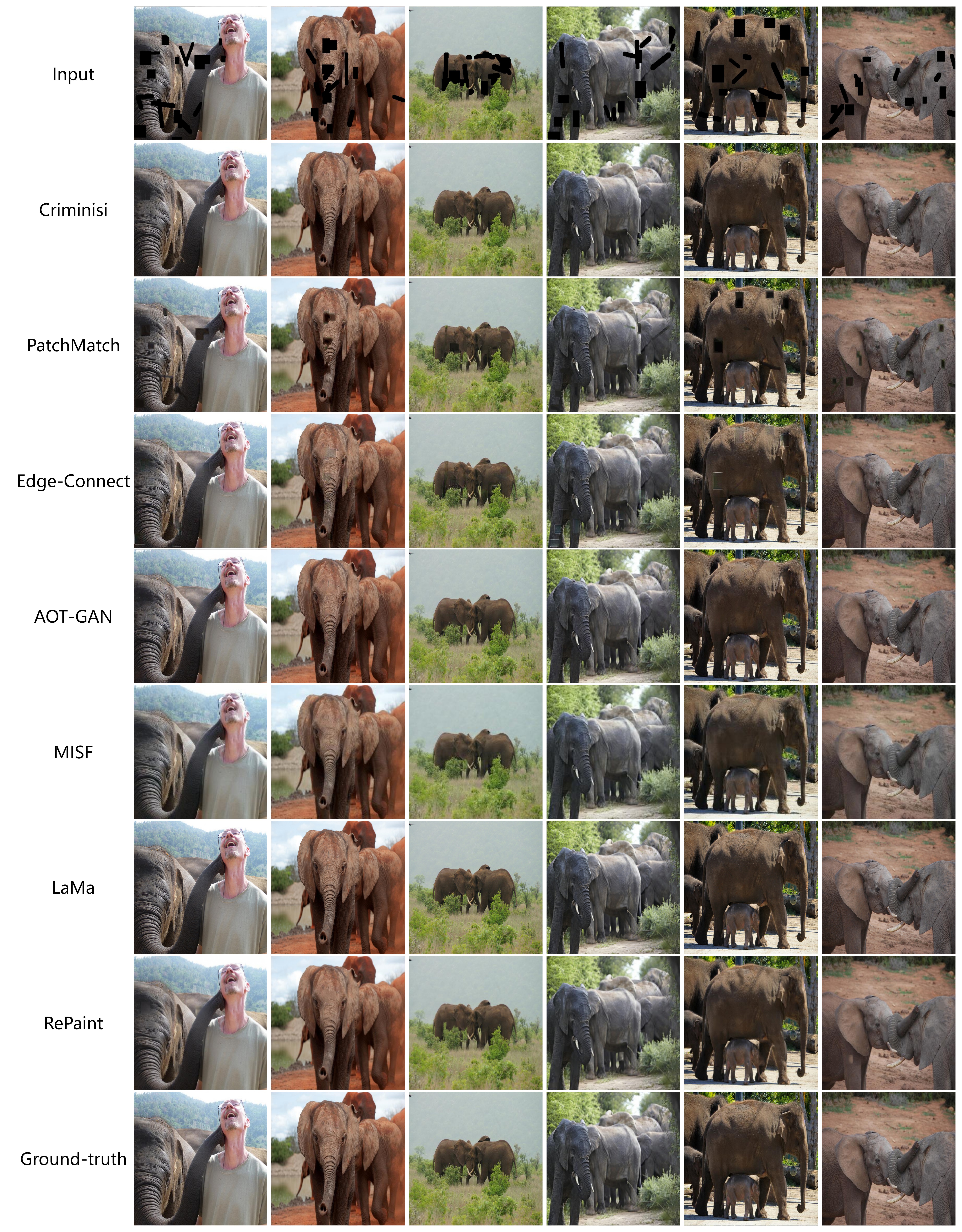}
\caption{Qualitative comparison of exemplar inpainting results (elephant dataset).}
\label{fig:fig11}
\end{figure}

\begin{figure}[htbp]
\centering
\includegraphics[width=1\linewidth]{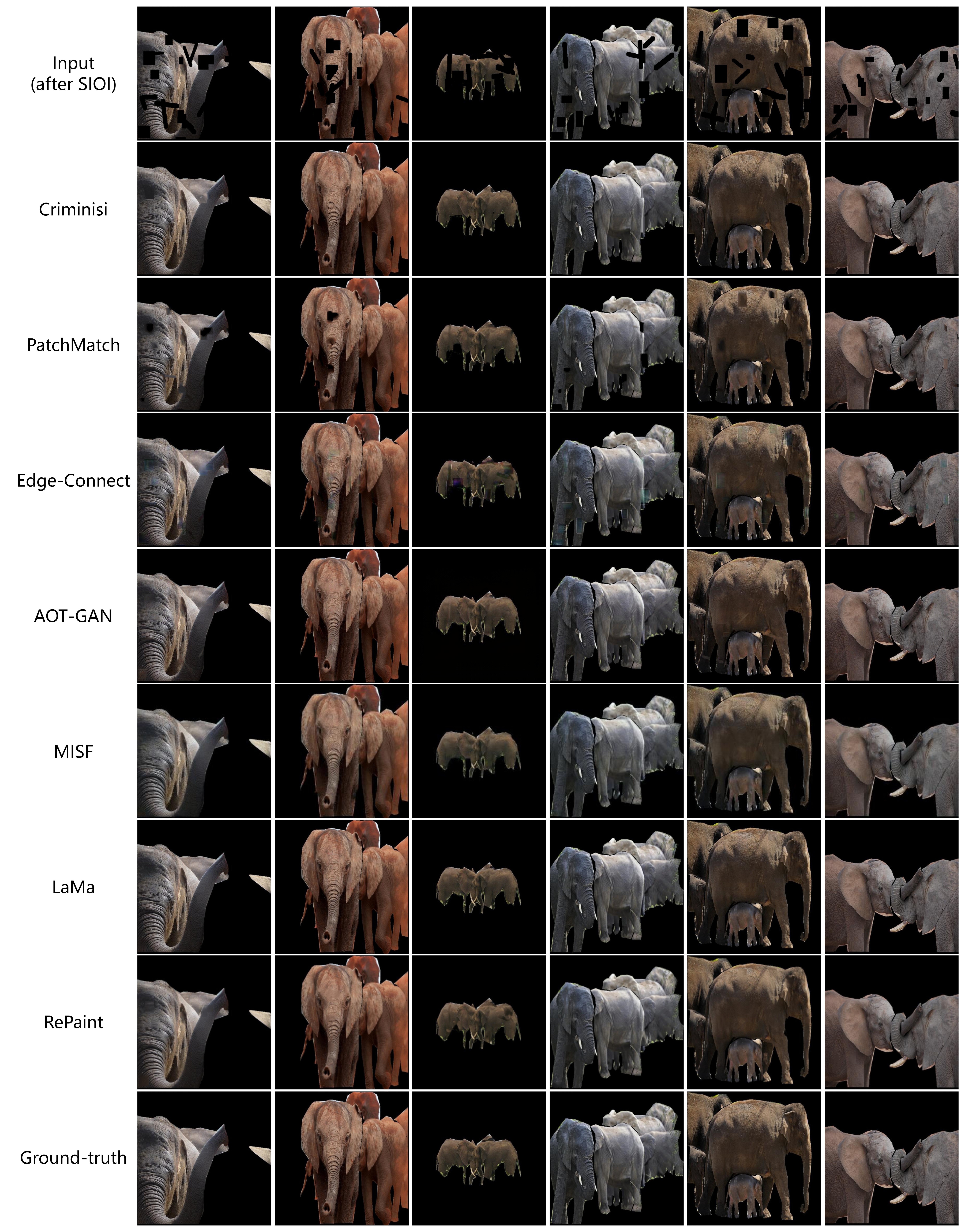}
\caption{Qualitative comparison of exemplar SIOI inpainting results (elephant dataset).}
\label{fig:fig12}
\end{figure}

\begin{figure}[htbp]
\centering
\includegraphics[width=1\linewidth]{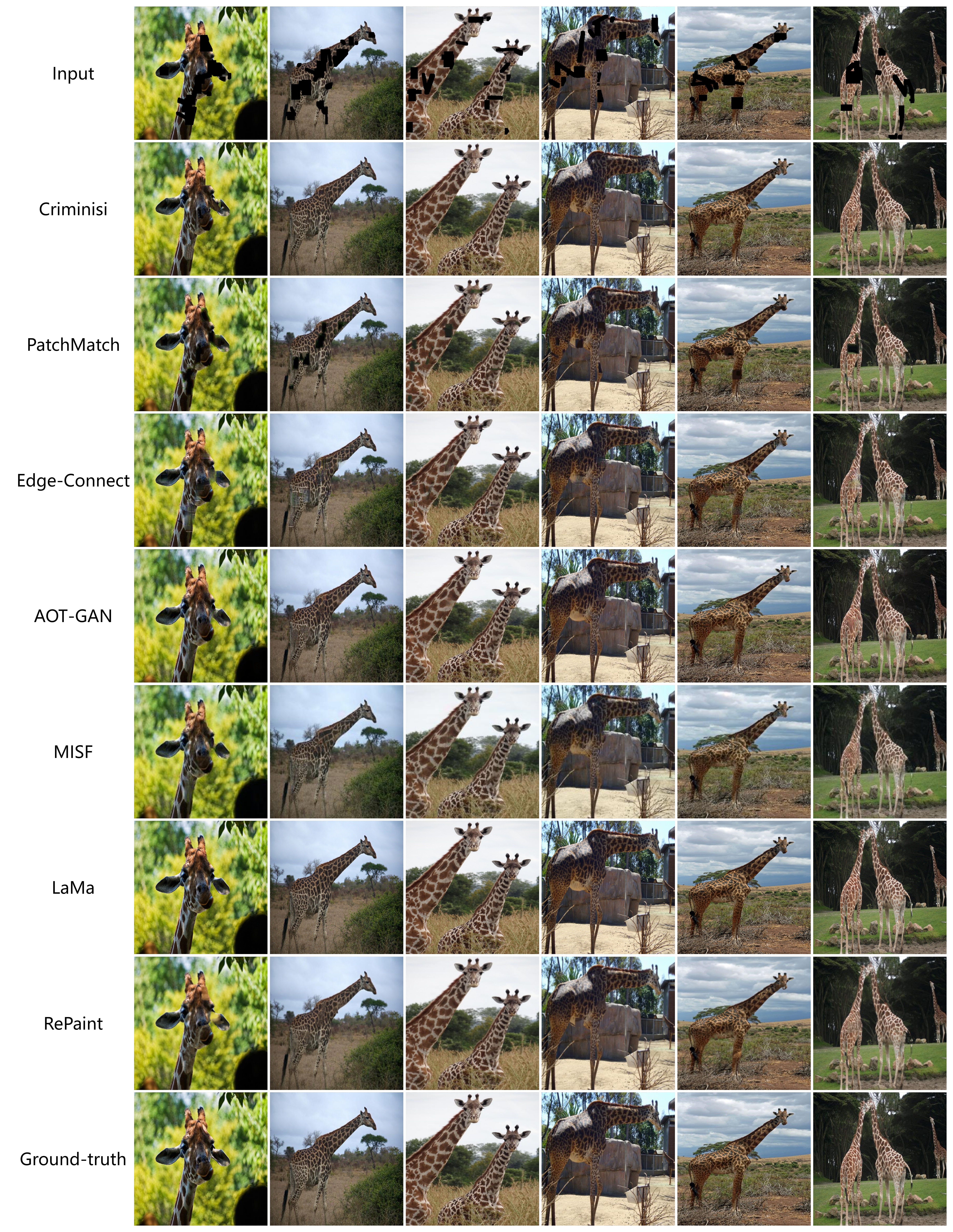}
\caption{Qualitative comparison of exemplar inpainting results (giraffe dataset).}
\label{fig:fig13}
\end{figure}

\begin{figure}[htbp]
\centering
\includegraphics[width=1\linewidth]{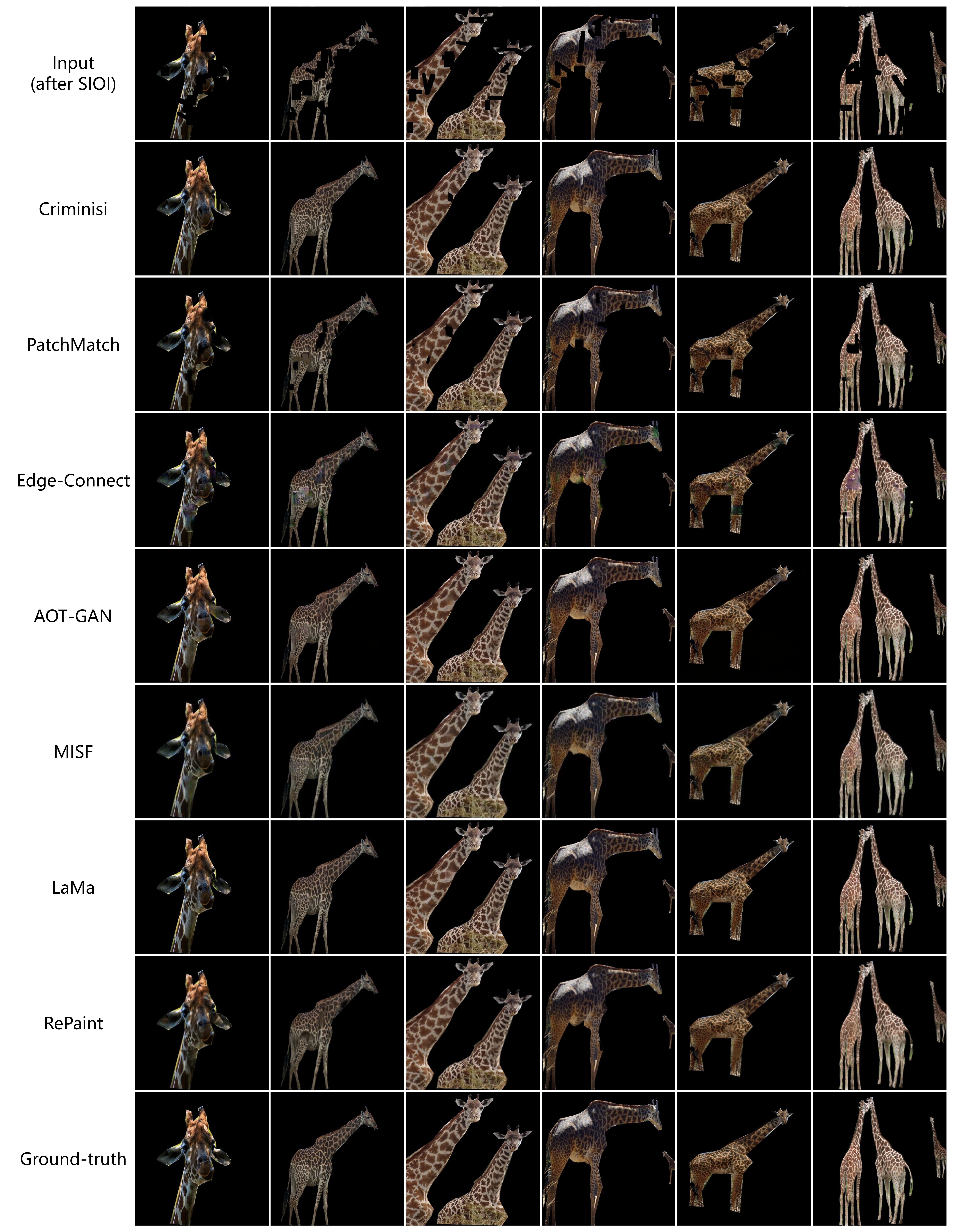}
\caption{Qualitative comparison of exemplar SIOI inpainting results (giraffe dataset).}
\label{fig:fig14}
\end{figure}

\begin{figure}[htbp]
\centering
\includegraphics[width=1\linewidth]{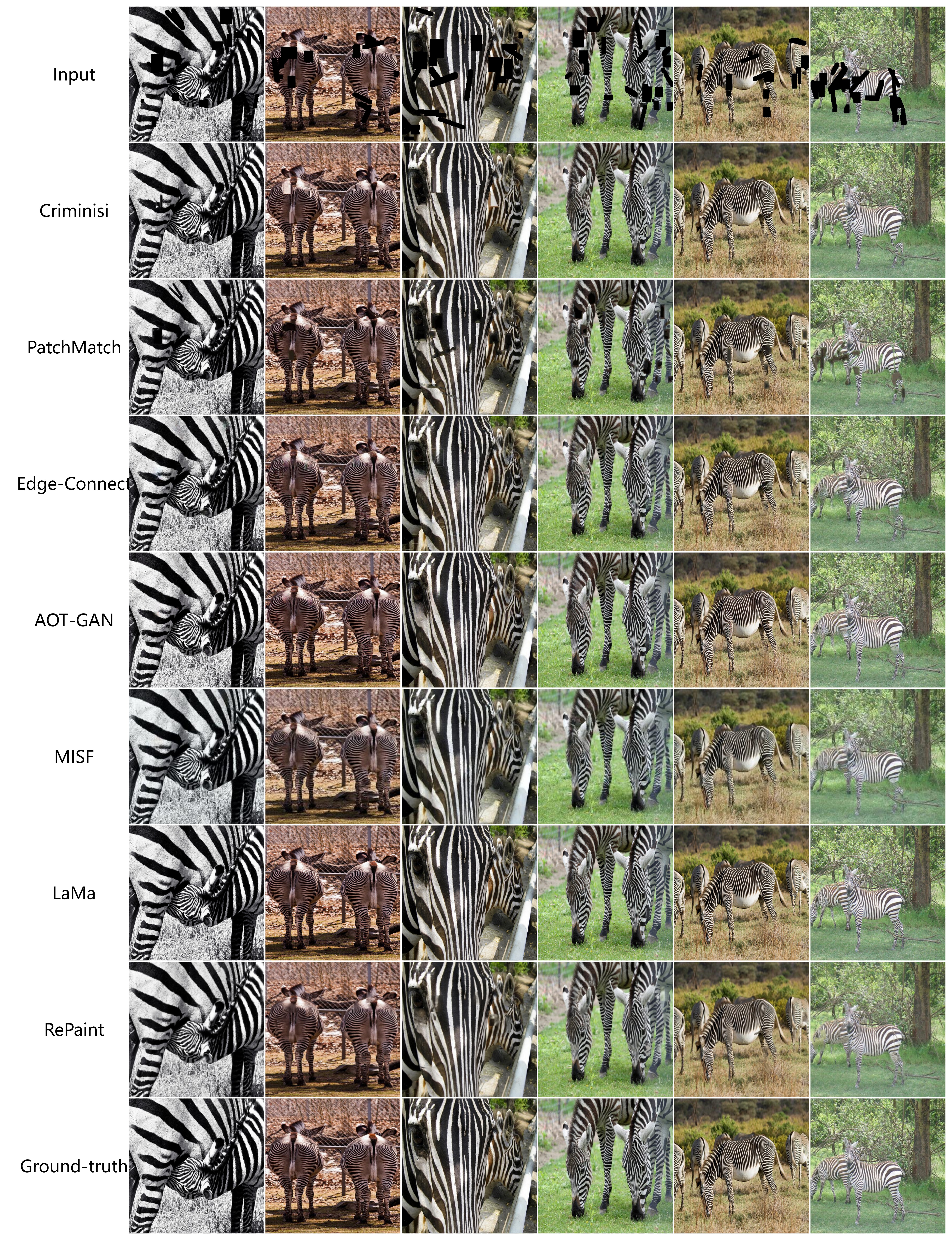}
\caption{Qualitative comparison of exemplar inpainting results (zebra dataset).}
\label{fig:fig15}
\end{figure}

\begin{figure}[htbp]
\centering
\includegraphics[width=1\linewidth]{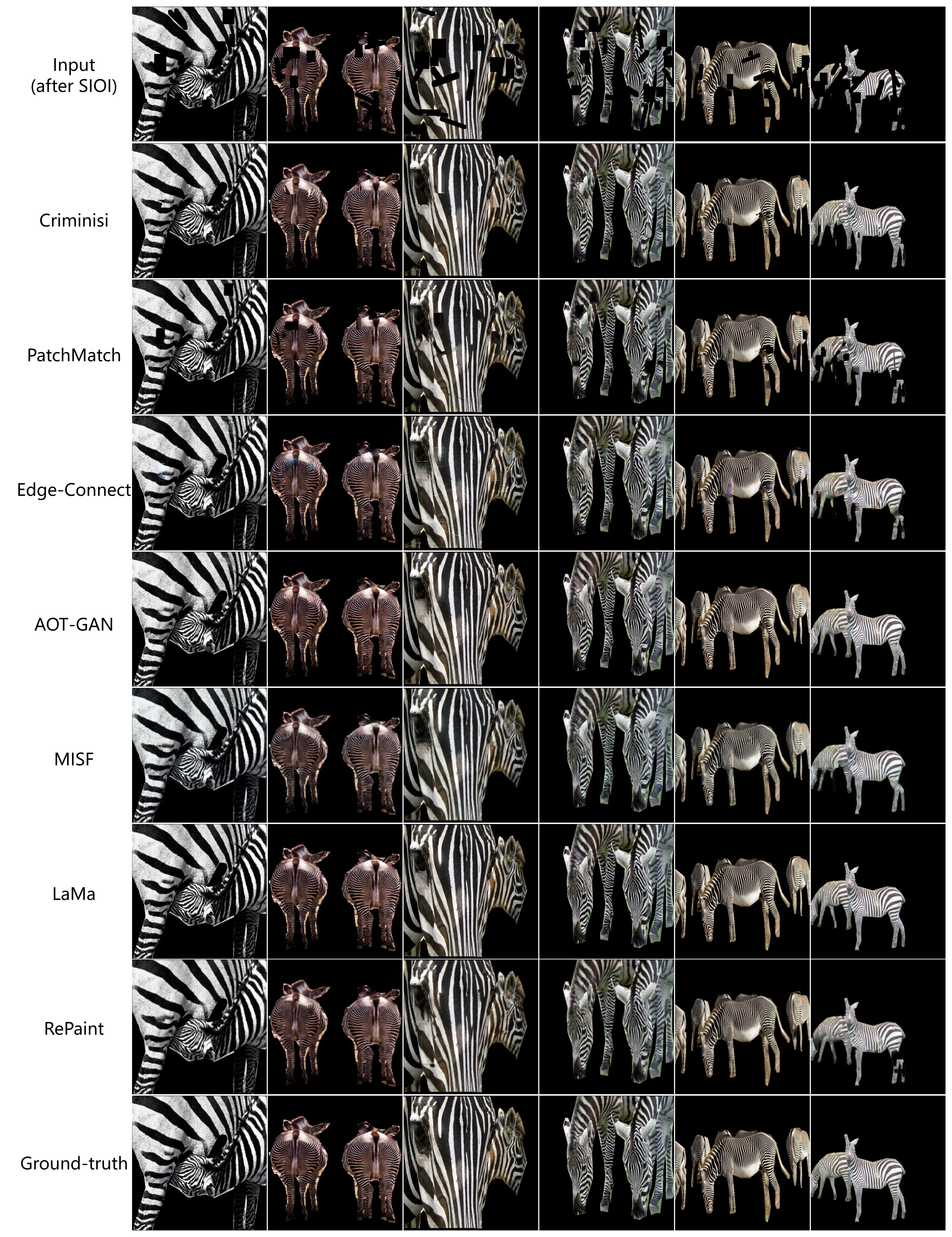}
\caption{Qualitative comparison of exemplar SIOI inpainting results (zebra dataset).}
\label{fig:fig16}
\end{figure}

Qualitative results corroborate the effectiveness of SIOI's as a preprocessing stage. By producing background-suppressed object-centric representations, SIOI enables conventional inpainting algorithms to recover more consistent and plausible structures. Figures~\ref{fig:fig9}--\ref{fig:fig16} illustrate characteristic failure modes of baseline methods: Criminisi~\citep{criminisi2004region} lacks semantic understanding, leading to structural discontinuities; PatchMatch~\citep{barnes2009patchmatch} exhibits visible patch artifacts; AOT-GAN~\citep{zeng2022aggregated} can introduce jagged patterns from dilated convolutions; Edge-Connect~\citep{nazeri2019edgeconnect} is sensitive to edge misalignment; MISF~\citep{li2022misf} tends to oversmooth edges and lighting; LaMa~\citep{suvorov2022resolution} sometimes oversmooths fine details; and RePaint~\citep{lugmayr2022repaint} may yield inconsistent textures. 

When preceded by SIOI, all methods generally show improved edge sharpness, greater texture consistency, and stronger semantic coherence. SIOI performs particularly well on structurally rigid objects (e.g., teapots). Performance on non-rigid classes (giraffes, zebras, elephants) is comparatively weaker, which we attribute to higher intra-class morphological variability; addressing this may require larger or more adaptive models and additional training data.

\subsection{Ablation studies}
\label{subsec:ablation}

We conducted ablation studies on the Teapot dataset under the same training conditions as the preliminary validation. For each ablation, we measured SSIM, PSNR, MAE, and LPIPS over five independent runs; means (and standard deviations where provided) are reported in Table~\ref{tab:tab4}. Key ablations:

\textbf{Ablation I (no Stage I)}: We removed the SIOI front-end while retaining object-centric input/output. Quantitatively, the metric differences were small (SSIM: 0.976 vs. 0.978; PSNR: 33.95 dB vs. 33.86 dB), but subjectively the reconstructions showed poorer edge localization and more jagged artifacts. This highlights that pixel-wise metrics may understate perceptual degradations.

\textbf{Ablation II (full-scene input)}: Replacing object-centric I/O with full-scene input caused substantial performance drops (SSIM: 0.964, PSNR: 29.43 dB, MAE: 2.09, LPIPS: 0.07). Visual inspection showed marked blurring and structural errors, which confirmed the advantage of an object-first processing paradigm.

\begin{figure}[htbp]
   \centering
    \includegraphics[width=1\linewidth]{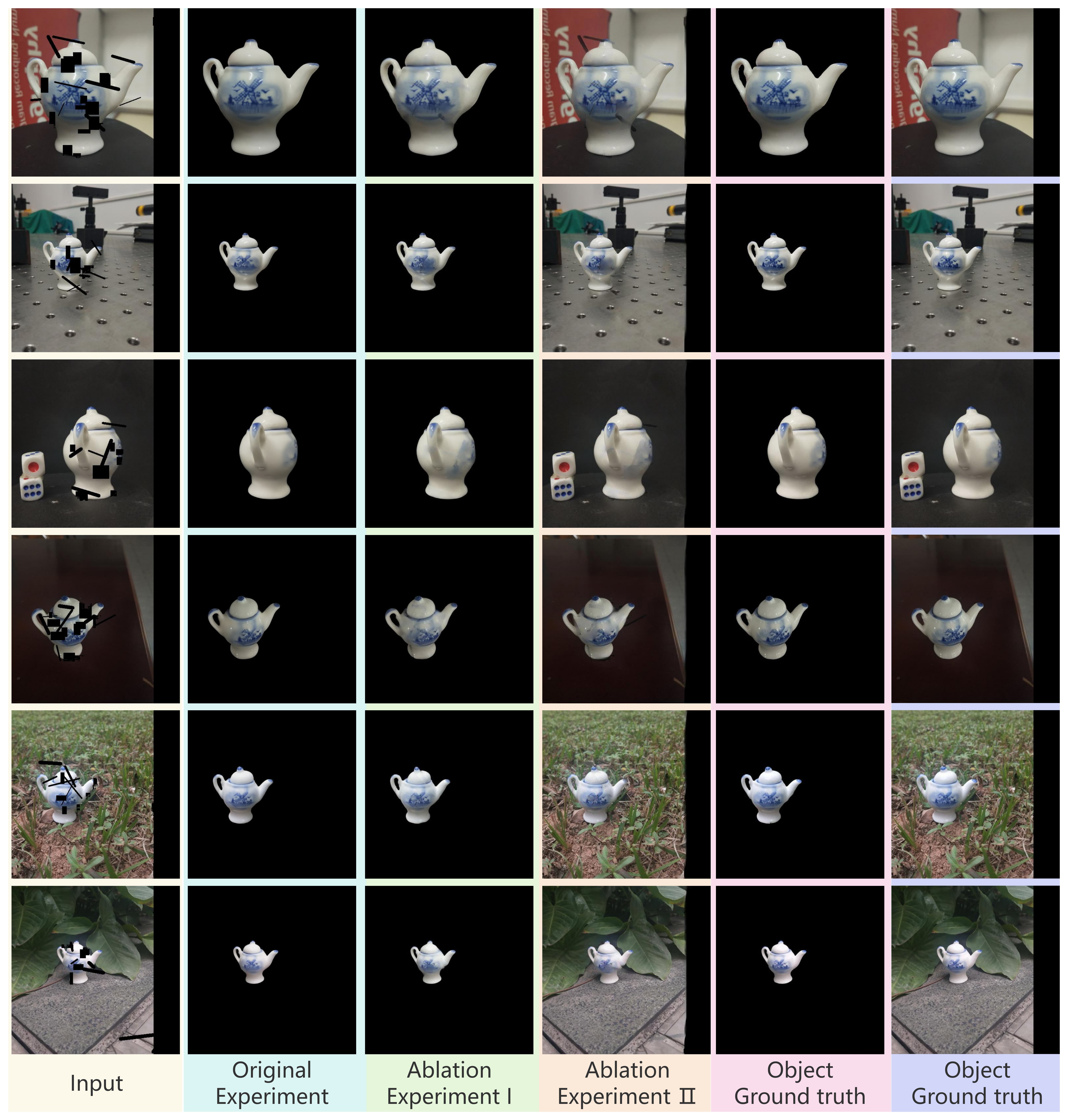}
    \caption{Illustration of ablation experiment.}
    \label{fig:fig17}
\end{figure}

Overall, the full pipeline (with SIOI) provides sharper edges, better semantic consistency, and more photorealistic textures than ablated variants, validating the two-stage design where object localization and structural priors precede texture synthesis.

\begin{table}[htbp]
\centering
\small
\caption{Ablation study results on the Teapot dataset (mean $\pm$ std over five runs). Arrows indicate preferred direction.}
\label{tab:tab4}
\begin{tabular}{lcccc}
\toprule
Model & SSIM $\uparrow$ & PSNR (dB) $\uparrow$ & MAE $\downarrow$ & LPIPS $\downarrow$ \\
\midrule
Full model (SIOI) & $0.98 \pm 0.01$ & $33.86 \pm 0.38$ & $1.61 \pm 0.08$ & $0.02 \pm 0.01$ \\
Ablation I (no Stage I) & $0.98 \pm 0.01$ & $33.95 \pm 0.19$ & $1.31 \pm 0.06$ & $0.02 \pm 0.01$ \\
Ablation II (full-scene I/O) & $0.96 \pm 0.01$ & $29.43 \pm 0.25$ & $2.09 \pm 0.15$ & $0.07 \pm 0.02$ \\
\bottomrule
\end{tabular}
\end{table}

\subsection{Robustness Analysis}
\label{subsec:robustness}

We further extend the preliminary validation on the Teapot dataset to evaluate SIOI's generalization under several challenging, out-of-distribution conditions. The scenarios tested are as follows:

\begin{itemize}
  \item \textit{Low illumination}: global brightness reduced by 50\%;
  \item \textit{Small targets}: objects that occupie $<10\%$ of the image area;
  \item \textit{Gaussian noise}: additive Gaussian noise with mean $\mu=0$ and standard deviation $\sigma=1$;
  \item \textit{Salt-and-pepper noise}: impulse noise with $p_{\mathrm{salt}}=0.01$ and $p_{\mathrm{pepper}}=0.01$;
  \item \textit{Multi-object occlusion}: three target instances present and occluded;
  \item \textit{Motion blur}: linear motion blur with a $15\times15$ kernel.
\end{itemize}

\begin{figure}[htbp]
   \centering
    \includegraphics[width=0.7\linewidth]{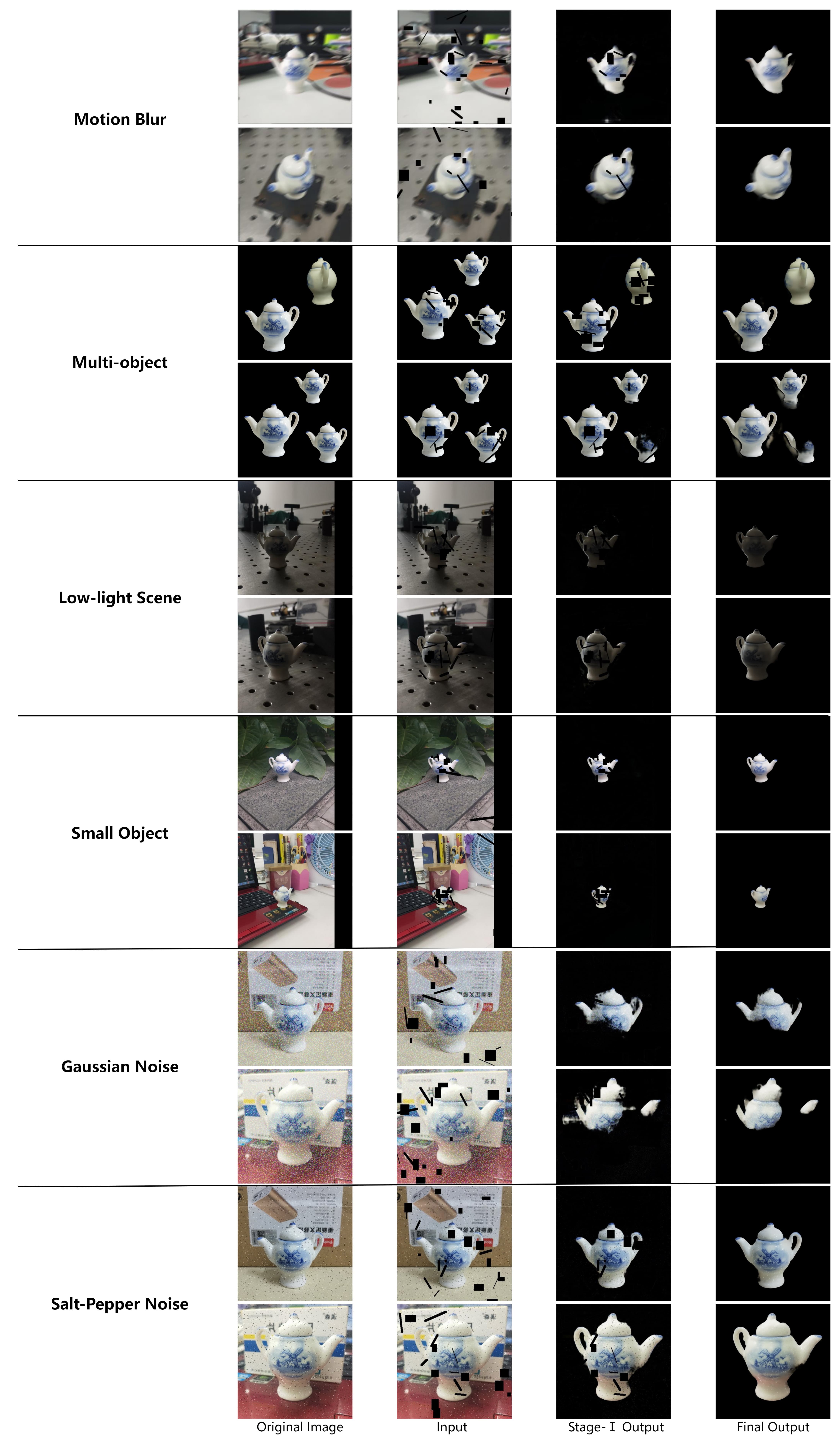}
    \caption{Representative examples from the robustness experiments (Teapot dataset).}
    \label{fig:fig18}
\end{figure}

These corruptions were not seen during training. For qualitative inspection, we selected two representative samples per condition; we did not perform a full quantitative study here due to the limited number of samples per corruption type (see Fig.~\ref{fig:fig18}).

Qualitatively, SIOI maintains satisfactory performance under reduced illumination, small targets, salt-and-pepper noise, and moderate motion blur. We attribute this robustness to two design choices: (1) spectral-domain processing (FFC) that reduces sensitivity to uniform brightness changes and some motion artifacts and (2) attention-inspired modules that amplify small, target-specific cues. In contrast, performance degrades under strong, high-entropy Gaussian noise: when noise overwhelms deterministic gradient and spectral cues, the SIOI front-end may fail to produce a stable object prior, which in turn affects downstream reconstruction.

Multi-object occlusion (three or more instances) also reduces performance, consistent with limited attentional capacity in both biological and artificial systems: competition among multiple salient inputs dilutes the module's selective gain, echoing the biased-competition account of attention \citep{desimone1995neural}.

These observations highlight two practical points: (i) SIOI is effective for a wide range of structured degradations but is vulnerable to non-structured, high-entropy noise; (ii) because SIOI acts as a preprocessor, its failures propagate downstream — therefore, designing fail-safes (e.g., detection of low-confidence priors, fallback full-scene processing, or ensembling) is important in deployment. Fig.~\ref{fig:fig18} presents representative qualitative examples under the tested conditions.

\section{Discussion}

Image inpainting remains an open problem with no universal solution. In this work, we propose Specific Object-of-Interest Imaging (SIOI), a bio-inspired front end that reorients the pipeline toward object-centric processing. Rather than attempting holistic, whole-image completion, SIOI first isolates and enhances the target object to provide a refined structural and semantic prior for any downstream inpainting network. Our experiments show that pre-prepending SIOI consistently improves the performance of a broad set of inpainting frameworks.

SIOI is inspired by the object-centric and attention-modulated processing of the ventral visual pathway: selective gain amplifies relevant features and suppresses background clutter (e.g. V4-mediated contour integration) \citep{roelfsema1998object,desimone1995neural}. Computationally, this reduces informational redundancy and mitigates error propagation from irrelevant background regions that often confound holistic methods (GAN-, transformer-, or diffusion-based).

Robustness experiments indicate that SIOI is resilient to many structured degradations (illumination changes, impulse noise, motion blur), but is relatively vulnerable to high-entropy, unstructured noise and to scenarios with strong multi-object competition. Because downstream results depend critically on SIOI's outputs, we recommend incorporating mechanisms to (i) detect low-confidence priors, (ii) enable graceful fallbacks (e.g., revert to whole-image strategies), or (iii) combine multiple priors (ensemble) to mitigate catastrophic failures.

\begin{figure}[htbp]
   \centering
    \includegraphics[width=1\linewidth]{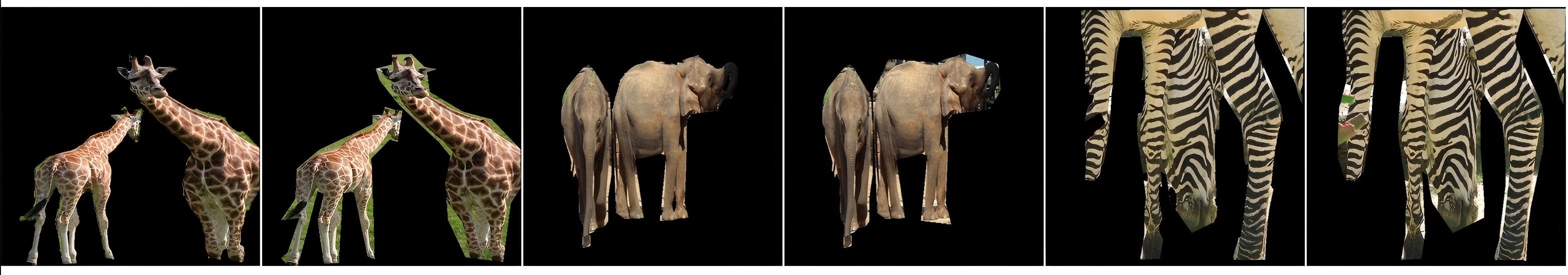}
    \caption{SIOI output exceeds the quality of its ground-truth annotations.}
    \label{fig:fig19}
\end{figure}

We observed that, in some cases, the SIOI output exceeds the quality of its ground-truth annotations, producing a bias in quantitative evaluations (see Fig.~\ref{fig:fig19}). This phenomenon mirrors the dorsal–ventral interaction in biological vision, where spatial localization informs object recognition, and underscores that improvements in object-level preprocessing constitute an important research frontier.

The inpainting network used in Stage II functions primarily as a proof-of-concept to validate the SIOI prior; therefore, it is not intended to represent an upper bound on performance. Future work may combine SIOI with more sophisticated inpainting architectures. The principal contribution of this study is the SIOI paradigm itself, a shift from direct holistic inpainting toward object-first prior generation.

We acknowledge limitations in the current generalization: because the present SIOI instantiation targets specific semantic categories, each distinct category effectively corresponds to a separate dataset and requires retraining of the object imaging network. However, the module can be specialized and scaled to support technology transfer and task-specific deployment. Extending SIOI to produce category-agnostic or dynamically reconfigurable priors (for example, via task-driven modulation analogous to prefrontal control) is an important direction for future research.

\section{Conclusion}

We presented Specific Object-of-Interest Imaging (SIOI), a lightweight, bio-inspired preprocessing paradigm for object-centric visual restoration. SIOI isolates and enhances the target object to produce a high-fidelity structural prior that improves downstream inpainting models. Empirical results across multiple datasets and baselines demonstrate that SIOI consistently elevates structural fidelity and perceptual quality, and that it is particularly effective for rigid, well-defined objects.

Limitations include sensitivity to high-entropy noise and reduced effectiveness under multi-object competition. Future work will focus on (i) improving robustness to unstructured noise (for example, through neuromorphic gain-control mechanisms or noise-aware spectral filtering), (ii) enabling task-driven, category-adaptive modulation, and (iii) validating the approach with perceptual user studies and neuroimaging experiments.

\appendix
\section{Extended mathematical notation and biological mapping}
\label{app:notation}

\begin{appendices}
\begin{table}[htbp]
\centering
\small
\caption{Extended mathematical notation and biological correlations.}
\label{tab:neuro_math_notation}
\footnotesize
\begin{tabular}{p{1.1cm} p{4.2cm} p{3.3cm} p{3.3cm}}
\toprule
\textbf{Symbol} & \textbf{Mathematical definition} & \textbf{Biological mechanism} & \textbf{Network implementation} \\
\midrule
$\Gamma$ & Target-sensitive measurement matrix: $\Gamma k \approx \sum_{m=1}^{M} (\Gamma \omega_m) \otimes k_m$ & LGN / V4 attentional filtering & SIOI spectral filter \\
\hline
$k$ & Scene tensor: $k \in \mathbb{R}^{H\times W\times C}$ & Retinal output & Input to SIOI (degraded image) \\
\hline
$\otimes$ & Diffeomorphic operator in feature space: $(\Gamma \omega_m)\otimes k_m$ & Foveal cortical magnification / feature warping & FFC-based feature warping \\
\hline
$\xi_j$ & Basis feature vector: $\xi_j\in\mathbb{R}^D$ & V1 simple-cell responses & Multi-scale encoder features \\
\hline
$\omega_m$ & Feature weights / attentional gains & Neural resource allocation & Learned attention weights \\
\hline
$k_i$ & Latent object tensor & IT-like object representation & Stage-I object prior \\
\hline
$\phi$ & Projection operator approximating $I-P_{\mathrm{bg}}$ & Background suppression & TRNet / projection module \\
\hline
$D_{\mathrm{KL}}$ & Kullback–Leibler divergence & Statistical coding efficiency & Loss/convergence diagnostics \\
\hline
$\mathcal{F}$ & Spectral feature map; e.g., $\mathcal{F}'=\mathrm{IFFT2}(\mathrm{Conv_{freq}}(\mathrm{FFT2}(\mathcal{F})))$ & Spatial-frequency processing (V1–V4) & FFCModule operations \\
\hline
$\mathcal{L}_{\mathrm{total}}$ & Composite loss: $\mathcal{L}_{\mathrm{total}}=\lambda_1\mathcal{L}_{\mathrm{recon}}+\lambda_2\mathcal{L}_{\mathrm{perc}}+\lambda_3\mathcal{L}_{\mathrm{style}}$ & Task-driven modulation & Phased training strategy \\
\bottomrule
\end{tabular}
\end{table}

\end{appendices}

\bibliographystyle{elsarticle-num}  
\bibliography{references}         

\end{document}